\documentclass[twocolumn]{aastex62}

\usepackage{amsmath}
\usepackage{color}
\usepackage{graphicx}
\usepackage[caption=false]{subfig}
\usepackage{tikz}
\usetikzlibrary{shapes.geometric, arrows}
\usepackage{multirow}
\usepackage{enumitem}
\usepackage{algpseudocode}

\DeclareUnicodeCharacter{2212}{-}

\newcommand{\Var}{{\text{Var}}}

\shorttitle{Lost Horizon}
\shortauthors{Bassett et al.}

\begin{document}

\title{Lost Horizon: Quantifying the Effect of Local Topography on Global 21-cm Cosmology Data Analysis}

\correspondingauthor{Neil Bassett}
\author{Neil Bassett}
\affiliation{Center for Astrophysics and Space Astronomy, Department of Astrophysical and Planetary Sciences, University of Colorado, Boulder, CO 80309, USA}

\author{David Rapetti}
\affiliation{NASA Ames Research Center, Moffett Field, CA 94035, USA}
\affiliation{Research Institute for Advanced Computer Science, Universities Space Research Association, Mountain View, CA 94043, USA}
\affiliation{Center for Astrophysics and Space Astronomy, Department of Astrophysical and Planetary Sciences, University of Colorado, Boulder, CO 80309, USA}

\author{Keith Tauscher}
\affiliation{Center for Astrophysics and Space Astronomy, Department of Astrophysical and Planetary Sciences, University of Colorado, Boulder, CO 80309, USA}

\author{Bang~D.~Nhan}
\affiliation{Department of Astronomy, University of Virginia, Charlottesville, VA 22903, USA}
\affiliation{National Radio Astronomy Observatory  (NRAO) Technology Center (NTC), Charlottesville, VA 22903, USA}

\author{David~D.~Bordenave}
\affiliation{Department of Astronomy, University of Virginia, Charlottesville, VA 22903, USA}
\affiliation{National Radio Astronomy Observatory  (NRAO) Technology Center (NTC), Charlottesville, VA 22903, USA}

\author{Joshua~J.~Hibbard}
\affiliation{Center for Astrophysics and Space Astronomy, Department of Astrophysical and Planetary Sciences, University of Colorado, Boulder, CO 80309, USA}

\author{Jack~O.~Burns}
\affiliation{Center for Astrophysics and Space Astronomy, Department of Astrophysical and Planetary Sciences, University of Colorado, Boulder, CO 80309, USA}


\email{Neil.Bassett@colorado.edu}

\begin{abstract}

We present an investigation of the horizon and its effect on global 21-cm observations and analysis. We find that the horizon cannot be ignored when modeling low frequency observations. Even if the sky and antenna beam are known exactly, forward models cannot fully describe the beam-weighted foreground component without accurate knowledge of the horizon. When fitting data to extract the 21-cm signal, a single time-averaged spectrum or independent multi-spectrum fits may be able to compensate for the bias imposed by the horizon. However, these types of fits lack constraining power on the 21-cm signal, leading to large uncertainties on the signal extraction, in some cases larger in magnitude than the 21-cm signal itself. A significant decrease in signal uncertainty can be achieved by performing multi-spectrum fits in which the spectra are modeled simultaneously with common parameters. The cost of this greatly increased constraining power, however, is that the time dependence of the horizon's effect, which is more complex than its spectral dependence, must be precisely modeled to achieve a good fit. To aid in modeling the horizon, we present an algorithm and Python package for calculating the horizon profile from a given observation site using elevation data. We also address several practical concerns such as pixelization error, uncertainty in the horizon profile, and foreground obstructions such as surrounding buildings and vegetation. We demonstrate that our training set-based analysis pipeline can account for all of these factors to model the horizon well enough to precisely extract the 21-cm signal from simulated observations.

\end{abstract}


\section{Introduction}
\label{Introduction}

The 21-cm spin-flip transition of neutral hydrogen provides a unique method of observing the intergalactic medium in the early universe. Thanks to the expansion of the universe, which redshifts the transition from its rest frequency of 1420 MHz to $\sim1 - 200$ MHz, the spectrum of the transition can be used to study the evolution of the universe during the Dark Ages as well as the history of the first astrophysical objects during Cosmic Dawn. The utility of the 21-cm transition as a cosmological probe ceases after the Epoch of Reionization 
(EoR), during which radiation from the first generation of sources ionizes nearly all of the hydrogen.

The global signal refers to the sky-averaged component of the redshifted 21-cm spectrum and is generally described by its brightness temperature, which is measured relative to the radio background (normally assumed to be the Cosmic Microwave Background).
The global signal is expected to have two absorption troughs at frequencies corresponding to the Dark Ages and Cosmic Dawn. Recently, the Experiment to Detect the Global EoR Signature (EDGES) reported an absorption trough in the range of the Cosmic Dawn portion of the spectrum that differs significantly in both shape and magnitude from predictions based on the standard cosmological model, $\Lambda$CDM, and theoretical models of the first generation of astrophysical objects.

Due to concerns surrounding possible unmodeled systematics in the data (see, e.g., \citealt{Hills:2018, Bradley:2019, Singh:2019, Sims&Pober:2020}) and the potential for the EDGES signal, if confirmed, to uncover non-standard physics (see, e.g., \citealt{Barkana:2018, Fialkov:2018, Ewall-Wice:2018, Feng&Holder:2019}), independent validation of the EDGES result is of the utmost importance. A number of other experiments are attempting to provide an independent measurement of the global signal. Ground-based instruments include the Radio Experiment for the Analysis of Cosmic Hydrogen (REACH; \citealt{deLeraAcedo:2019}), the Shaped Antenna measurement of the background RAdio Spectrum (SARAS; \citealt{Patra:2013,Singh:2017}), Sonda Cosmol\'ogica de las Islas para la Detecci\'on de Hidr\'ogeno Neutro (SCI-HI; \citealt{Voytek:2014}), the Large-aperture Experiment to detect the Dark Ages (LEDA; \citealt{Price:2018}), the Broadband Instrument for Global HydrOgen ReioNization Signal (BIGHORNS; \citealt{Sokolowski:2015}), the Cosmic Twilight Polarimeter (CTP; \citealt{Nhan:2019}), the Mapper of the IGM Spin Temperature (MIST)\footnote{\url{http://www.physics.mcgill.ca/mist/}}, and Probing Radio Intensity at high-z from Marion (PRIzM; \citealt{Philip:2019}). A lunar-based instrument, the Dark Ages Polarimeter PathfindER (DAPPER; \citealt{Burns:2021a, Burns:2021b}), is also under development.

Perhaps the most difficult obstacle in global 21-cm observations is separating the 21-cm signal from bright foreground emission, which dwarfs the cosmological signal by four to six orders of magnitude \citep{Furlanetto:2006, Mozdzen:2019}. While the foreground emission is composed primarily of synchrotron radiation, which can be described by a power law in frequency, the total foreground is not a single power law but the average of many different power laws with different spectral indices. In addition, the beam chromaticity inherent to real instruments making wide-band measurements, in combination with the spatially anisotropic nature of the foreground, introduces further spectral structure. Along with instrumental effects, the ionosphere can also introduce chromatic distortions that do not cancel out over time \citep{Shen:2021}.

The customary method for global 21-cm analysis is to attempt to ``correct'' observations for antenna beam systematics (see, e.g., the beam chromaticity factor of \citealt{Bowman:2018}) before using a polynomial-based model to fit the foreground. However, this parameterization of the foreground is insufficient to model the response of realistic, chromatic beams \citep{Hibbard:2020} and may, in fact, lead to the recovery of false signals \citep{Tauscher:2020b}.

An alternative to polynomial-based models is forward modeling, which employs simulated observations to create models with which to fit data. One such forward modeling-based approach is to create training sets composed of many variations in the systematics, including the intrinsic foreground, the beam, and the instrument receiver. The training sets are then decomposed into basis vectors with which to build a model through Singular Value Decomposition (SVD). This framework is employed by our data analysis pipeline, which is laid out in a series of four papers: \citealt{paperI} (Paper I), \citealt{paperII} (Paper II), \citealt{paperIII} (Paper III), and \citealt{paperIV} (Paper IV). The \texttt{pylinex} code,\footnote{\url{https://bitbucket.org/ktausch/pylinex/}} which supports the pipeline, is publicly available \citep{pylinex:21}. In addition to our pipeline, others have adopted other forward-modeling based approaches for global signal extraction (see, e.g., \citealt{Anstey:2020}, which describes a framework based on a Bayesian nested sampling algorithm).

In this paper we will discuss an element of low frequency observations that has heretofore been overlooked by 21-cm global signal experiments: the horizon. Specifically, we will identify the consequences of a ``lost'' horizon; that is when the horizon used for modeling the foreground and signal components differs from reality. We begin in Section \ref{effect_of_horizon} by performing simulations of the beam-weighted foreground component and examining the bias imposed by a lost horizon on observed spectra. In Section \ref{calculating_horizon}, we present an algorithm for calculating the horizon profile at any location on a spherical body so that the presence of the horizon can be incorporated in the forward modeling of observations. Section \ref{accounting_for_horizon} then discusses how our training set-based pipeline can account for the presence of the horizon such that the 21-cm signal can be extracted accurately. Finally, we conclude by discussing the implications of this work for the planning and analysis of low frequency experiments.

\section{Quantifying the Effect of a Horizon}
\label{effect_of_horizon}

\begin{figure*}
    \centering
    \includegraphics[width=\textwidth]{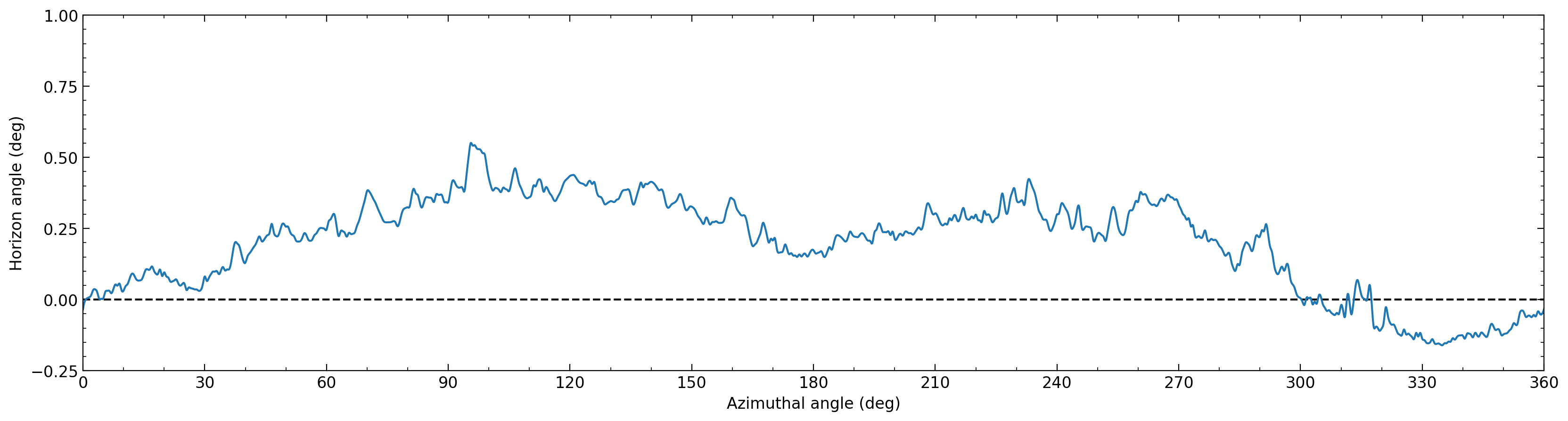}
    \caption{Profile of the horizon as seen from the location of the EDGES instrument calculated from elevation data. $0^{\circ}$ azimuth corresponds to North with the angle increasing clockwise. The dashed black line shows a flat horizon for reference. For details about the calculation see Appendix \ref{horizon_derivation}.}
    \label{fig:EDGES_horizon}
\end{figure*}

In this section, we discuss how ignoring or misrepresenting the shape of the horizon may negatively affect the extraction of the 21-cm signal. The horizon at a given location may be composed of several different features, such as the natural terrain (i.e. hills or mountains), vegetation, or human-made obstructions. While these features may influence the propagation of low frequency radio waves in different ways, in this section we will consider only the case in which incoming radiation is perfectly blocked at or below the horizon. This example will serve to illustrate the importance of the horizon even in its most simplified form, while further complications are addressed in Section \ref{accounting_for_horizon}.

\subsection{Methodology}

It is important to note that the antenna beam is inextricably linked to the horizon, and will thus influence all low frequency observations. If the beam were sufficiently small, we could effectively ignore the horizon as the sensitivity of the antenna would be so low near the horizon as to be negligible. However, for the antennas used in global 21-cm experiments, this will almost certainly never be the case.
In order to simulate a sufficiently general experiment, we will use a Gaussian beam with a frequency-dependent Full Width at Half Maximum (FWHM), $\alpha(\nu)$, given by
\begin{equation}
\label{beam_equation}
    B(\nu, \theta, \phi) \propto \exp \Bigg\{-4\ln 2 \bigg[\frac{\theta}{\alpha(\nu)}\bigg]^2\Bigg\}.
\end{equation}
The particular FWHM used in all of the simulations in this paper is quadratic in frequency, following
\begin{equation}
\label{beam_fwhm_equation}
    \alpha(\nu) = 115^{\circ} - 0.3^{\circ} \Big(\frac{\nu}{1\textup{ MHz}}\Big) + 0.001^{\circ} \Big(\frac{\nu}{1\textup{ MHz}}\Big)^2.
\end{equation}

When performing simulated signal extractions, we will compare the effect of the horizon on several different analysis methods. First, we will consider a single spectrum fit, in which all data are averaged together into one spectrum. This is the method employed by the EDGES team in \cite{Bowman:2018}. We will also consider multi-spectrum fits, which may utilize spectra that are binned in time, as well as full Stokes polarization measurements.

Ignoring receiver systematics for now (which are addressed in Paper IV), the spectrum observed by an antenna will be composed of two primary components: the 21-cm signal and the beam-weighted foreground. Under the assumption of a large beam, the horizon will modify the measured spectrum of both of these components. Due to its isotropic nature, the consequence of the horizon for the 21-cm signal is simple to model. We will use a spatially- and spectrally-dependent quantity $\mathcal{H}$ to describe the effect of the horizon on incoming radiation, where a value of 1 means the incoming radiation is unattenuated in that direction (e.g. far above the horizon), 0 means the incoming radiation is completely blocked (e.g. far below the horizon), and intermediate values indicate that the incoming radiation is partially attenuated. Assuming this definition of $\mathcal{H}$, the observed global 21-cm spectrum will be attenuated by a factor $\lambda(\nu)$ given by
\begin{equation}
    \lambda(\nu) = \frac{\int_{4\pi} B(\nu, \theta, \phi)\mathcal{H}(\nu, \theta, \phi)d\Omega}{\int_{4\pi} B(\nu, \theta, \phi)d\Omega}.
\end{equation}
Note that in some cases the horizon $\mathcal{H}$ may be frequency dependent, which we will discuss further in Section \ref{accounting_for_horizon}.

While the multiplicative attenuation factor imposed on the signal may be relatively simple, the exact form of the observed beam-weighted foreground spectrum is complicated by the anisotropic foreground emission. The synchrotron radiation that composes the majority of the foreground emission is concentrated in the plane of the Galaxy. Thus, when the plane of the Galaxy is near the horizon, we should expect the effect of the horizon on the foreground spectrum to be more pronounced. Section \ref{beam_weighted_foreground_bias} will explore how the horizon alters the measured foreground spectrum in more detail.

\subsection{Beam-Weighted Foreground Bias}
\label{beam_weighted_foreground_bias}

\begin{figure*}[t]
    \centering
    \subfloat{\includegraphics[width=0.33\textwidth]{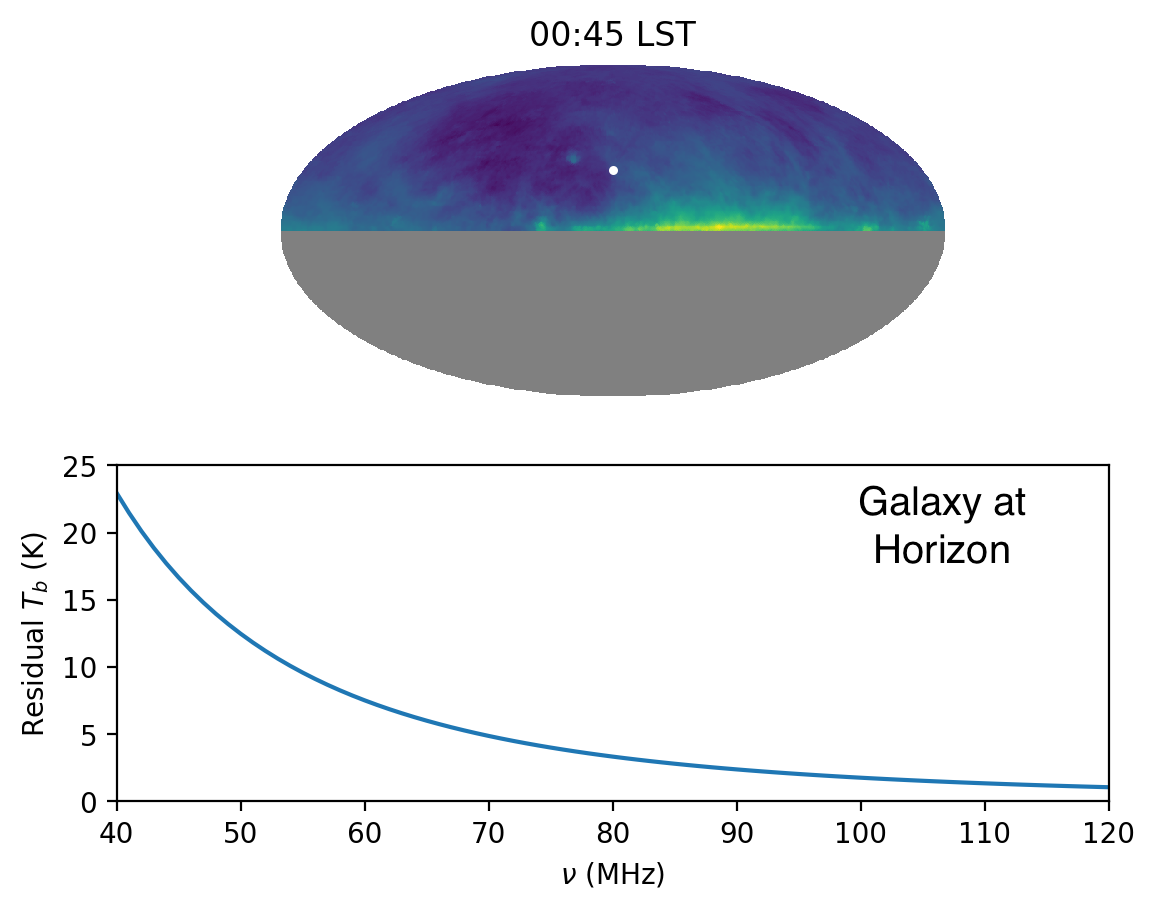}}
    \subfloat{\includegraphics[width=0.33\textwidth]{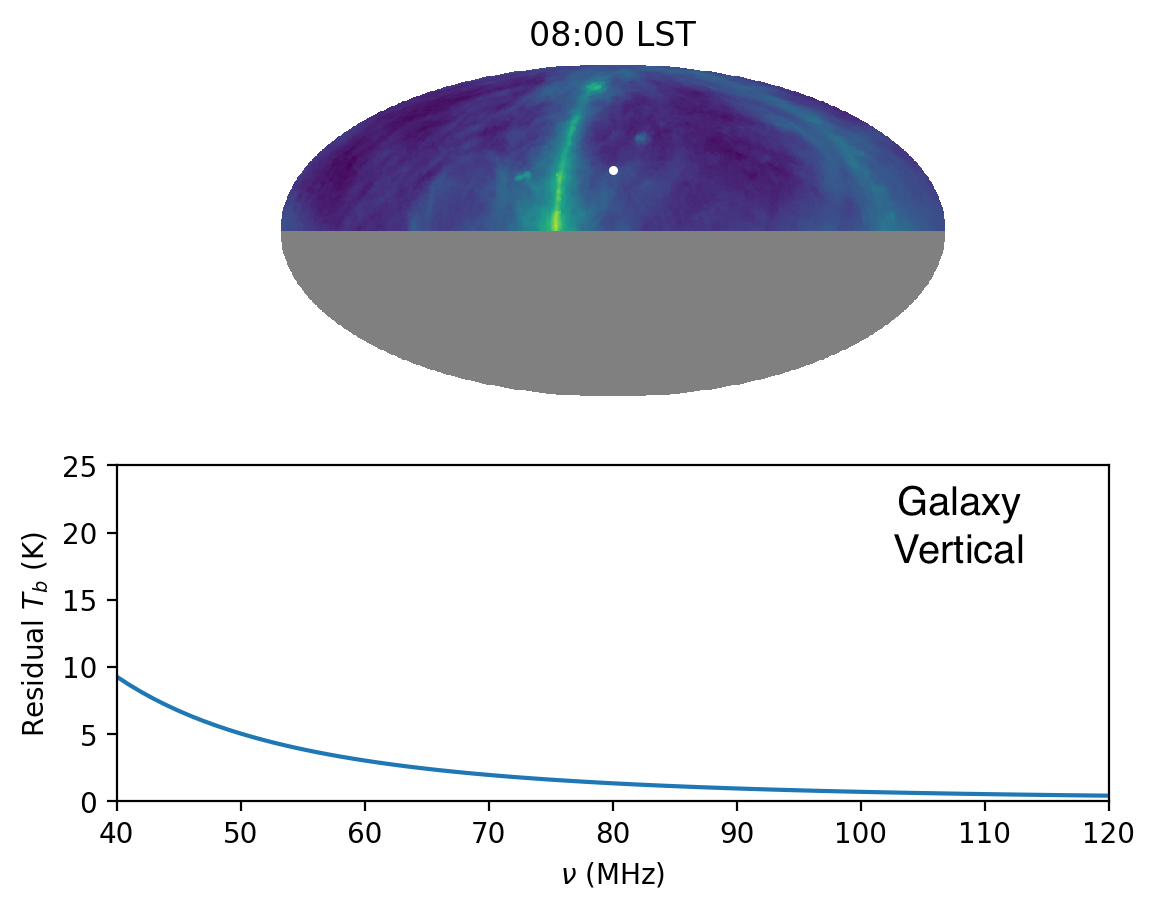}}
    \subfloat{\includegraphics[width=0.33\textwidth]{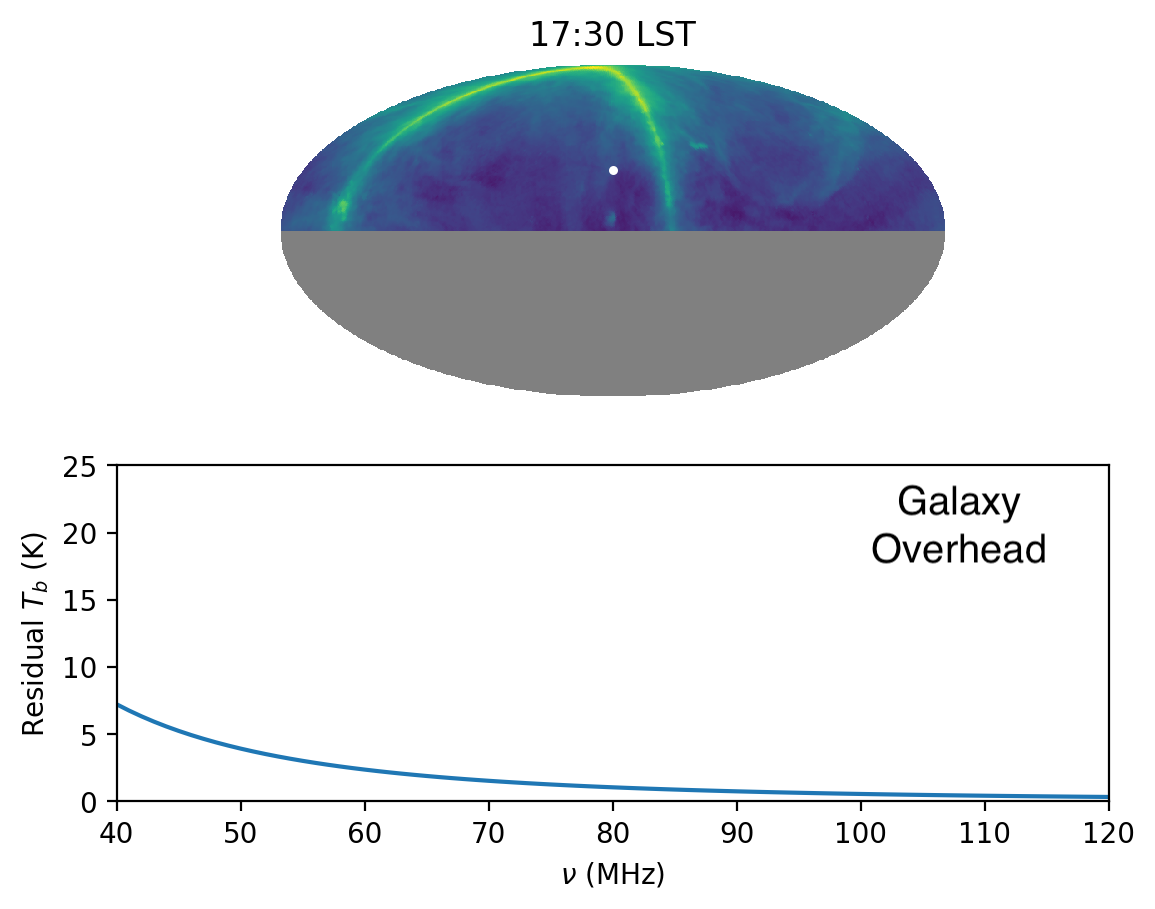}}
    \caption{\textit{Top}: Mollweide projection of the sky at three different LSTs as seen from the location of the EDGES experiment in Western Australia. The white dot indicates the location of the South Celestial Pole around which the Galaxy appears to revolve. Grey indicates portions of the sky below the flat horizon. The three LSTs were chosen such that the plane of the Galaxy is either approximately parallel to the horizon, approximately perpendicular to the horizon, or largely above the horizon. \textit{Bottom}: Residuals in the Stokes I beam-weighted foreground spectrum when using a perfectly flat horizon (black dashed curve in Figure \ref{fig:EDGES_horizon}) compared to using  the EDGES horizon derived from elevation data (blue curve in Figure \ref{fig:EDGES_horizon}) for each of the three LSTs. The spectra are calculated assuming the sky emission is described by the Haslam 408 MHz map \citep{Haslam:1982, Remazeilles:2015} extrapolated to lower frequencies through a power law with a constant spectral index of -2.5.}
    \label{fig:EDGES_horizon_galaxy_position_and_residuals}
\end{figure*}

To begin, we consider simulations of observations performed from the location of the EDGES experiment (\href{https://goo.gl/maps/u8s7owniWV8TYUK38}{$26^{\circ}\ 42^{\prime}\ 53.8^{\prime\prime}$ S, $116^{\circ}\ 36^{\prime}\ 12.5^{\prime\prime}$ E}) in the Murchison Radio-astronomy Observatory in Western Australia. While the features of the surrounding terrain are quite small, the horizon as viewed from the EDGES location is not perfectly flat, as shown in Figure \ref{fig:EDGES_horizon}. For further details regarding the calculation of the horizon from elevation data, see Appendix \ref{horizon_derivation}.

\begin{figure}
    \centering
    \includegraphics[width=\columnwidth]{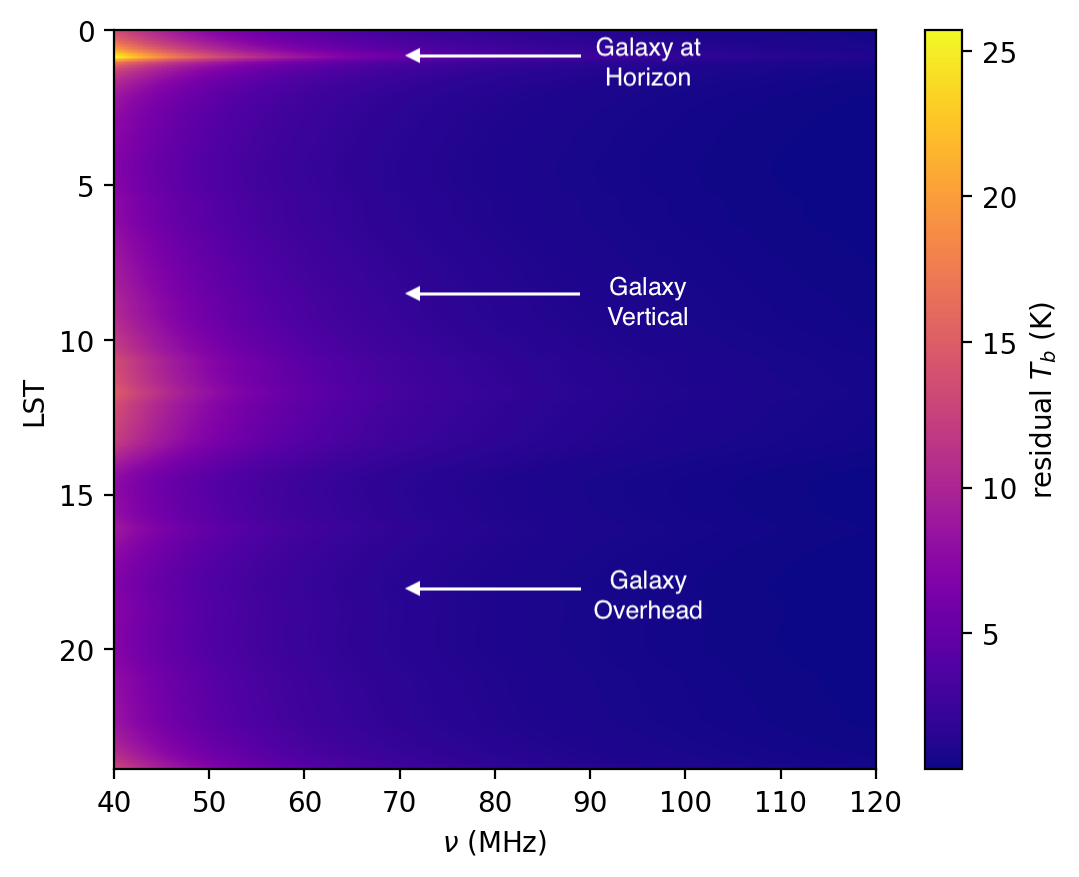}
    \caption{Waterfall plot showing the residual Stokes I beam-weighted foreground spectrum when using a flat horizon compared to the EDGES horizon, the same quantity as is plotted in the bottom panels of Figure \ref{fig:EDGES_horizon_galaxy_position_and_residuals}. The arrows and annotations point out the LSTs of the three Galaxy positions shown in the frames of Figure \ref{fig:EDGES_horizon_galaxy_position_and_residuals}. Although the spectral dependence at constant LST of this residual is simple and power law-like, the anisotropy of the galactic foreground introduces a complicated time dependence.}
    \label{fig:EDGES_waterfall}
\end{figure}

\begin{figure*}
    \centering
    \includegraphics[width=\textwidth]{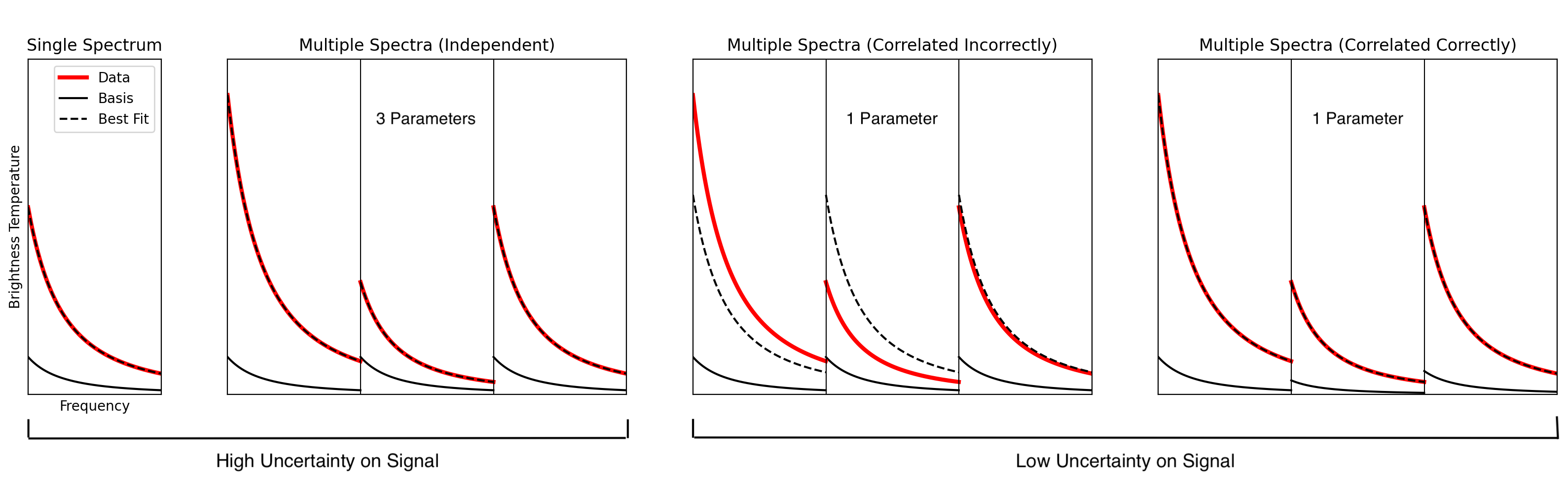}
    \caption{Schematic of how a time dependent change in magnitude can lead to problems when fitting correlated spectra simultaneously, even when the basis includes the correct spectral shape. The horizontal axis in all panels represents frequency while the vertical axis represents brightness temperature. All of the curves shown in this figure are power laws with the same exponent. The red curves represent data realizations to be fit with the basis vectors shown in black. The best fit in each case is indicated by a dashed curve. The left panel shows a single spectrum fit, the left middle panel shows three spectra that are fit independently, the right middle panel shows a fit to three correlated spectra in which the correlation is not represented correctly in the basis vector, and the right panel shows a fit to three correlated spectra when the correlations are properly described by the basis. Although the left and left middle panels display good fits, fits to single or multiple independent spectra have significantly less constraining power for global 21-cm signal extraction than fitting correlated spectra. While in the left middle panel three model parameters were fit to the data, only one was used in the right panel, thus providing tighter constraints thanks to the additional modeling of the correlations between spectra.}
    \label{fig:horizon_spectra_schematic}
\end{figure*}

As mentioned above, the effect of the horizon will be greater in magnitude when the plane of the Galaxy is lower in the sky. The top panels of Figure \ref{fig:EDGES_horizon_galaxy_position_and_residuals} show three different galactic orientations at various local sidereal times (LST). The residuals in the bottom panels quantify the difference in the measured beam-weighted foreground at one instant in time between the EDGES horizon and a flat horizon. While there is a large difference in the foreground spectra regardless of galactic position (note that the scale of the residuals is in Kelvin), the residual is maximized when the Galaxy is near $\theta \sim 90^{\circ}$ as in the frame on the left. It is important to note that the terrain surrounding the EDGES location is largely flat and the difference between the EDGES horizon and a flat horizon is small. These residuals, which are significantly larger in magnitude than the global 21-cm signal, are induced by a difference in the horizon of less than half a degree. While the exact form of the residual will depend on the horizon and the beam, both of which are specific to a given experiment, Figure \ref{fig:EDGES_horizon_galaxy_position_and_residuals} shows that even a small difference in the horizon can produce changes in the measured beam-weighted foreground at the Kelvin level and greater.

While it is clear that the horizon has a non-negligible effect on the measured foreground, it is important to determine how well this effect can be described by a given foreground model. Examining the bottom panels of Figure \ref{fig:EDGES_horizon_galaxy_position_and_residuals}, the spectral shape of the residuals appears at a glance to be smooth and power law-like. Given the relative simplicity of the spectral shape of the residuals, one might assume that the horizon does not pose a problem for global signal extraction at all as any bias imposed by the horizon will be absorbed by the foreground model. While this assumption may hold for the analysis of a single time-averaged spectrum, it breaks down when multiple spectra are fit simultaneously, as will be shown below.

To illustrate why the horizon cannot be disregarded in all but the simplest analysis, we examine the time dependence of the horizon's effect on the beam-weighted foreground spectrum in more detail. The magnitude of the residual as a function of sidereal time is shown more explicitly in the waterfall plot of Figure \ref{fig:EDGES_waterfall}. As noted above, the spectral shape of the residual always appears smooth and power law-like in frequency (along the horizontal axis). However, the shape of the residual in time (along the vertical axis) is quite complex, owing to the complicated spatial structure of both the EDGES horizon and the galactic emission. This complex time dependence of the horizon effect is crucial for precision global signal extraction, which requires modeling of the full waterfall plot image instead of individual spectra.

Consider the effect that a time-dependent bias has on the analysis of data that are binned in LST into three different spectra. A simplified schematic of this circumstance is shown in Figure \ref{fig:horizon_spectra_schematic}. In the schematic, the ``data'' and the ``basis'' are always composed of a single power law with the same exponent (i.e. spectral index). However, the magnitude of the power law varies across the three ``data'' spectra (red lines) in a similar fashion to how the horizon imposes a bias whose magnitude may vary across spectra that are binned in time. Two possibilities exist for how to carry out the analysis of the data binned into three spectra: the spectra can be treated independently or the spectra can be correlated. If the spectra are treated independently, then each spectrum gets its own separate basis vector and the coefficients multiplying these basis vectors may vary when performing a linear fit. In this case (shown in the left middle panel), the data spectra are fit well, as seen in the best fit curves (dashed lines) that overlap the data spectra exactly. On the other hand, when the spectra are correlated, there is only one basis vector that spans all three spectra and, consequently, there is only one coefficient determined by the fit. 
The right two panels show two different correlated fits. In the right middle panel, the basis vector is composed of three equal-magnitude power laws, implicitly assuming that the magnitude of the data is the same in all three spectra. As a result of this flawed assumption, the fit to the data is poor. However, in the right-most panel, we see that when the difference in magnitude between the spectra (i.e. the correlation between spectra) is encoded in the basis vector via time-varying amplitudes, the data are again fit well, this time with only one parameter.

Examining these four different cases, one might suppose that performing a single spectrum fit or fitting multiple spectra independently is the simplest and thus optimal method of global signal extraction. However, when we consider the level of constraint that these methods provide on the 21-cm signal, we find that they are not adequate. As demonstrated in Paper III, in order to realize the increased constraining power of fitting multiple spectra, the models used in the fit must include correlations between spectra (as in the right-most panel of Figure~\ref{fig:horizon_spectra_schematic}) in order to fully utilize the fact that, unlike the foreground, the signal has no time dependence. Through simulations, it was found that single spectrum or independent spectra fits produced RMS uncertainties at the several K level, while fitting multiple correlated time-binned spectra yielded uncertainties at the $\lesssim$ 10 mK level. For the global 21-cm signal, which almost certainly has a magnitude of $< 1$ K, fitting the spectra independently does not yield meaningful constraints on the signal. While in Figure \ref{fig:horizon_spectra_schematic}, we suppose that the different spectra represent different LST bins, the horizon will also alter the correlations between the Stokes polarization components. Thus, the lesson of Figure \ref{fig:horizon_spectra_schematic}, that the horizon must be carefully accounted for when analyzing multiple correlated spectra, holds true whether the multiple spectra represent polarization measurements, time bins, or both.

As discussed above, the horizon presents a trade off: the horizon can be largely ignored when performing fits that produce large uncertainties, such as fitting a single spectrum or multiple independent spectra (represented in the left two panels of Figure \ref{fig:horizon_spectra_schematic}), but for correlated multi-spectrum fits (represented in the right two panels of the same figure) that are able to precisely extract the global signal, the horizon must be treated in a detailed manner. However, properly modeling the effect of the horizon for correlated multi-spectrum fits poses an added difficulty, as illustrated by the waterfall plot shown in Figure \ref{fig:EDGES_waterfall}. When data are averaged to a single spectrum, the waterfall plot is collapsed along the vertical axis, essentially eliminating the complex time dependence and leaving only a relatively simple function of frequency. On the other hand, a fit to spectra that are binned in time must fit the full 2D structure of the waterfall plot. Further, if polarization measurements are included (see, e.g., \citealt{Nhan:2019}), the space that must be fit expands into a 3D space of four separate waterfall plots, one for each of the Stokes polarization parameters. While modeling the horizon may introduce additional complexity, it is imperative for performing correlated multi-spectrum fits that produce precise constraints on the 21-cm signal. In Section \ref{accounting_for_horizon} we will thus focus on how to utilize our pipeline to properly account for the horizon. However, it is important to note here that the horizon problem is not specific to our pipeline and will be relevant for any analysis method that relies on forward modeling.

\section{Determining the Horizon Profile}
\label{calculating_horizon}

Before addressing how to rigorously account for the horizon in more detail, we first discuss how to determine the shape of the horizon profile. Appendix \ref{horizon_derivation} presents a detailed derivation for calculating the horizon profile at any location on a spherical body using elevation data. A pseudo-code algorithm for performing this calculation is introduced in Appendix \ref{algorithm_pseudocode_appendix}. An implementation of the algorithm is made publicly available through the Simulating Horizon Angle Profile from Elevation Sets (SHAPES) code.\footnote{\url{https://github.com/npbassett/shapes}} The \texttt{shapes} code provides the ability to calculate the horizon for most locations on both the Earth and the Moon (for further details see Appendix \ref{horizon_profile_code}). A sample of horizons calculated with the \texttt{shapes} code at the locations of several different low frequency experiments is shown in Appendix \ref{horizon_comparison}.

\begin{figure*}
    \centering
    \includegraphics[width=\textwidth]{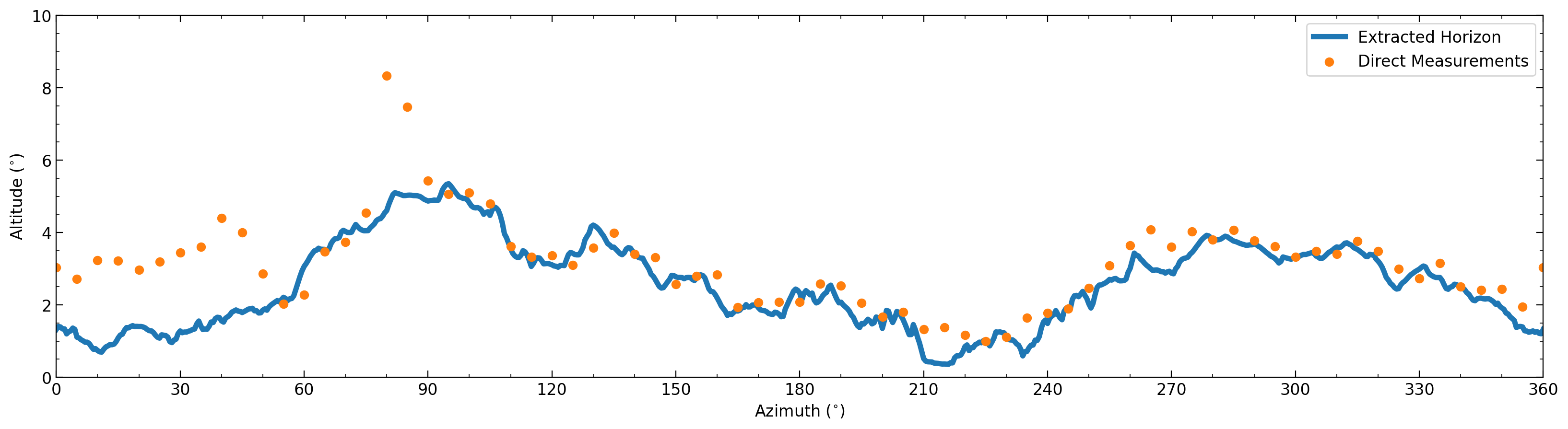}
    \caption{The horizon profile as seen from the location of the CTP instrument at Green Bank Observatory. $0^{\circ}$ azimuth corresponds to North with the angle increasing clockwise. The blue line was calculated from elevation data. The orange points are direct measurements made at the site. The locations at which the two profiles differ significantly can be attributed to trees and other structures not captured by elevation maps obstructing the sky.}
    \label{fig:greenbank_horizon}
\end{figure*}

\begin{figure}
    \centering
    \includegraphics[width=\columnwidth]{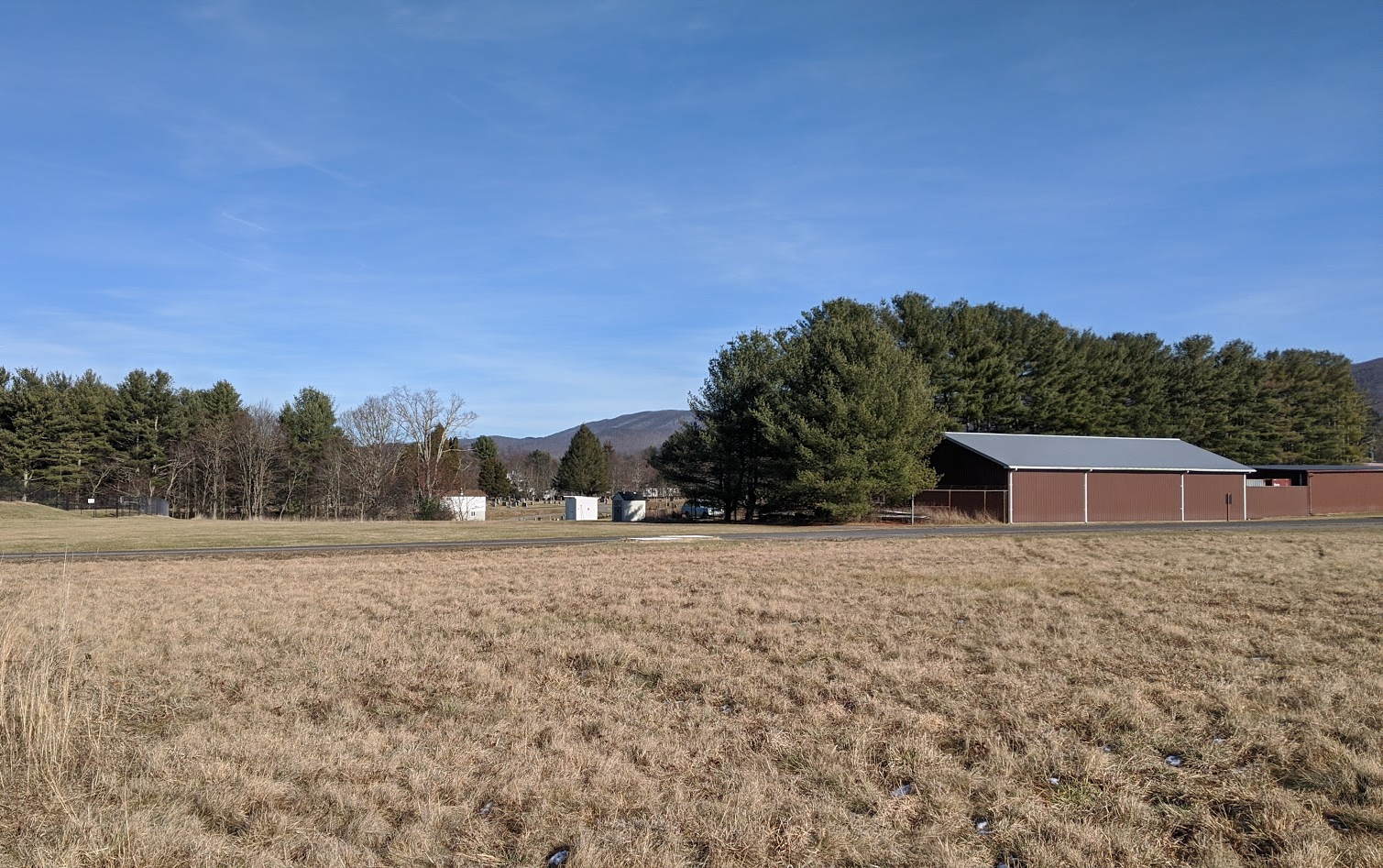}
    \caption{Photo looking approximately east-northeast ($\sim 70^{\circ}$ azimuth) from the location and on the same day that the direct measurements shown in Figure \ref{fig:greenbank_horizon} were made. The photograph's field of view is roughly $60^{\circ}$ in azimuth.}
    \label{fig:greenbank_east}
\end{figure}

\subsection{Comparison to Direct Measurement}
\label{comparison_to_measurement}

The Green Bank Observatory (GBO) is located in the National Radio Quiet Zone and is the location of the Cosmic Twilight Polarimeter (CTP), a proof-of-concept experiment to use projection-induced polarization measurements to constrain foreground emission to aid in extracting the global 21-cm signal \citep{Nhan:2019}.

During the Fall of 2019, measurements of the horizon from the site of the CTP instrument (\href{https://goo.gl/maps/eERmBVrQcRujUc3Q6}{$38^{\circ}\ 26^{\prime}\ 1.4^{\prime\prime}$ N, $79^{\circ}\ 49^{\prime}\ 22.9^{\prime\prime}$ W}) were performed at $5^{\circ}$ increments in azimuth with an analog surveying transit level, or theodolite. The direct measurements were rounded to the nearest arcminute. Figure \ref{fig:greenbank_horizon} compares these direct measurements to a horizon profile calculated with the \texttt{shapes} code using the algorithm derived in Appendix \ref{horizon_derivation}.


A close inspection of Figure \ref{fig:greenbank_horizon} reveals several portions of the horizon where the two profiles differ. These disagreements are mainly due to the presence of surrounding vegetation and building structures, which are not resolved in the elevation data from \texttt{shapes}. While the direct horizon measurements follow the top of the treeline, the radar used to perform the SRTM elevation measurements likely penetrates some distance into the canopy. The measurements were performed on February 11-22, 2000 during the flight of Space Shuttle Endeavor, meaning that there would not have been leaves on the surrounding deciduous trees, increasing the depth that the radar could penetrate into the canopy. In addition, if a patch of trees falls between two SRTM data points (which are $\sim 30$ m apart), the trees would be excluded from the horizon profile. Finally, in order to avoid numerical errors resulting in spuriously high horizon angles, a minimum angular distance of $\gamma = 0.001^{\circ}$ was imposed, excluding trees and any other objects within $\sim 100$ m. The presence of foreground obstructions explains why the direct measurements are always above the extracted profile in the locations where the two profiles differ significantly.

Figure \ref{fig:greenbank_east} shows one example of trees affecting the direct horizon measurements, but not the extracted horizon. The bank of trees on the right side of the photo accounts for the peak seen in the direct horizon measurements around $\sim 80^{\circ}$ azimuth. The trees on the left side of the photo correspond to the right side of the feature seen between $\sim 0 - 50^{\circ}$ azimuth in the direct measurements. Accounting for the presence of vegetation when performing signal extraction is discussed in Section \ref{vegetation}.

\subsection{Propagation of Error and Other Uncertainties}
\label{uncertainty_in_horizon_profile}

Error in the elevation measurements or the latitude and longitude coordinates of the elevation grid will produce an error in the calculated horizon profile. For a full derivation of the variance in the horizon angle $\eta$, see Appendix \ref{horizon_error_appendix}.

We also note that any deviation from perfect sphericity, such as the oblateness of the Earth, would, in the current version of the \texttt{shapes} code, introduce a small error into the horizon profile calculation due to (in the case of oblateness) differences in curvature between the equator and the poles. The oblateness of a spheroidal object can be described by the flattening factor $f = \frac{a - b}{a}$, where $a$ and $b$ are the semi-major and semi-minor axes, respectively. The flattening factor is only $f = 0.003$ for the Earth and $f = 0.001$ for the Moon, meaning that, for our purposes, oblateness is unlikely to affect the horizon significantly for each of these bodies.

Finally, diffraction effects may also introduce small deviations in the path of incoming radiation, making visible some sources that are slightly below the horizon. Although diffraction will not be explored further in this work, its effect can in principle be treated by altering the horizon map $\mathcal{H}$ used in performing simulations.

\section{Accounting for the horizon}
\label{accounting_for_horizon}

After demonstrating the importance of the horizon for modeling the foreground in Section \ref{effect_of_horizon} and presenting an algorithm for calculating the horizon in Section \ref{calculating_horizon}, we now discuss how to properly account for the presence of the horizon with our \texttt{pylinex} pipeline.

\subsection{Numerical Concerns}
\label{numerical_concerns}

We begin by addressing a relatively simple but important point regarding numerical errors caused by discretization. In reality, the brightness temperature measured by an antenna $T_{\textup{ant}}$ is given by an integral over the continuous functions of the sky temperature $T_{\textup{sky}}$, the horizon, and the beam of the antenna:
\begin{equation}
\label{continuous_foreground_equation}
    T_{\textup{ant}}(\nu) = \frac{\int_{4\pi}T_{\textup{sky}}(\nu,\theta,\phi)\mathcal{H}(\nu,\theta,\phi)B(\nu,\theta,\phi)d\Omega}{\int_{4\pi}B(\nu,\theta,\phi)d\Omega}.
\end{equation}
However, in order to produce simulations of the beam-weighted foreground, we must approximate the continuous integral with a finite sum. In the \texttt{HEALPix}\footnote{\url{https://healpix.sourceforge.io/}} format \citep{Gorski:2005}, each pixel of a map covers the same surface area on the sky so the discretized version of Equation \ref{continuous_foreground_equation} is then
\begin{equation}
\label{discrete_foreground_equation}
    T_{\textup{ant}}(\nu) = \frac{\sum_i^{N_{\textup{pix}}} T_{\textup{sky}}(\nu, i)\mathcal{H}(\nu, i)B(\nu,i)}{\sum_i^{N_{\textup{pix}}}B(\nu,i)},
\end{equation}
where $N_{\textup{pix}}$ is the total number of pixels in the map.

While necessary for producing simulations, this pixelization can introduce numerical errors that are exacerbated when including a horizon. The number of pixels in a \texttt{HEALPix} map is determined by the $N_{\textup{side}}$ parameter such that $N_{\textup{pix}} = 12(N_{\textup{side}})^2$, where $N_{\textup{side}}$ must be a power of $2$. When utilizing \texttt{HEALPix} maps to calculate the beam-weighted foreground with a horizon, care must be taken to ensure that the horizon is adequately represented in the discrete pixelization.

\begin{figure}
    \centering
    \includegraphics[width=\columnwidth]{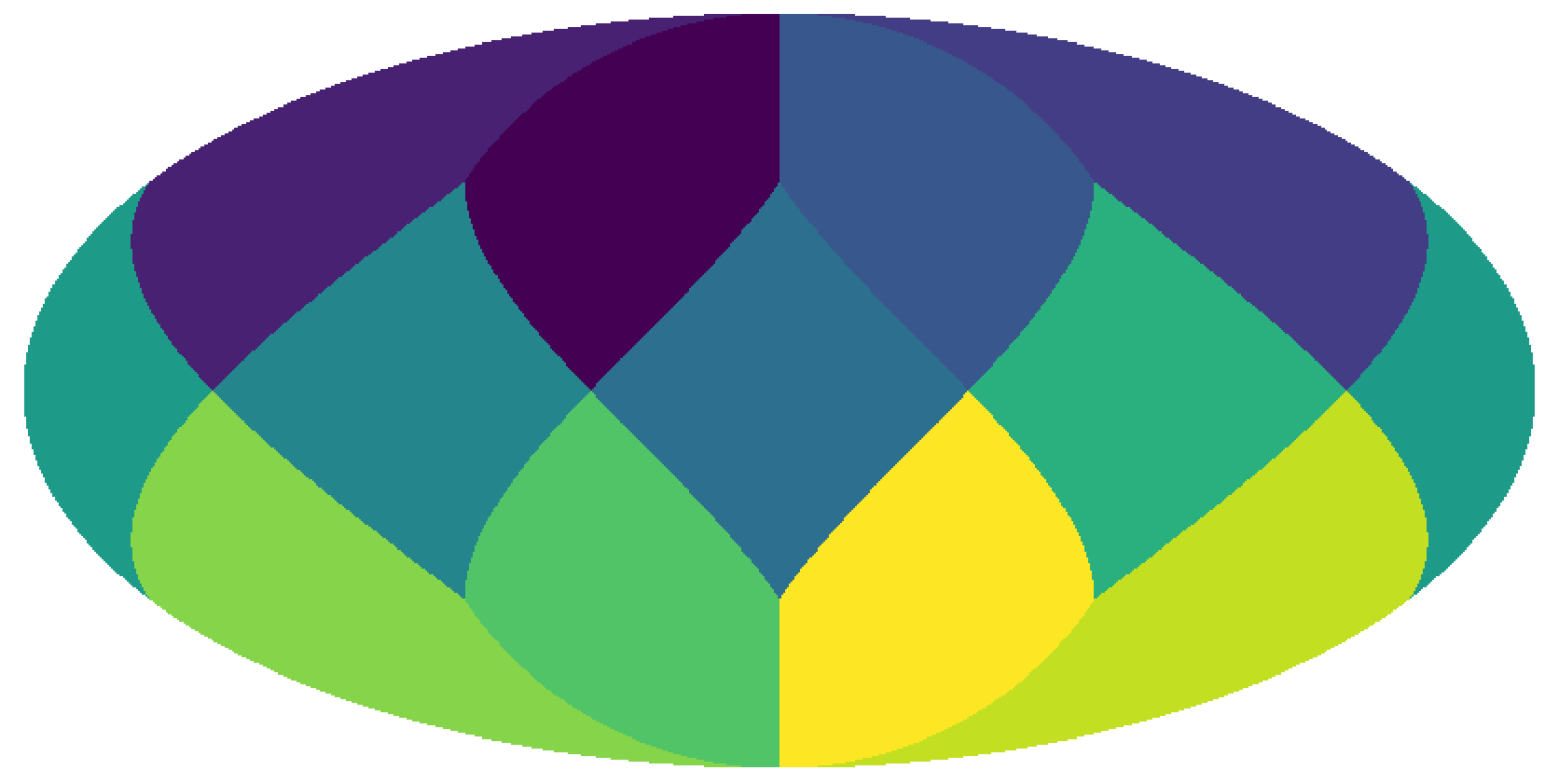}
    \caption{Mollweide projection of a \texttt{HEALPix} map with $N_{\textup{side}} = 1$. Colors were chosen to distinguish adjacent pixels.}
    \label{fig:healpix_base_map}
\end{figure}

Consider the lowest possible resolution \texttt{HEALPix} map shown in Figure \ref{fig:healpix_base_map}. Suppose that we would like to impose a perfectly flat horizon at $\theta = 90^{\circ}$ by setting pixels above the horizon to a value of 1 and pixels below the horizon to 0. The \texttt{HEALPix} format poses an obstacle because there exists a row of pixels centered on $\theta = 90^{\circ}$. This will always be true regardless of the value of $N_{\textup{side}}$. Thus, if we choose to set all of the pixels in the row to a value of 1, the horizon will block less than 50\% of the sky. Table \ref{tab:horizon_nside} shows a rough estimate of the error introduced to the beam-weighted foreground by this effect.

\begin{table}
    \centering
    \begin{tabular}{cccc}
    \hline
    \multirow{2}{*}{$N_{\textup{side}}$} & Blockage & Temperature & Map\\
    & Fraction & Residual (K) & Size (GB)\\
    \hline\hline
    1 & 0.33333 & 833 & 9.60e-8\\
    2 & 0.41667 & 417 & 3.84e-7\\
    4 & 0.45833 & 208 & 1.54e-6\\
    8 & 0.47917 & 104 & 6.14e-6\\
    16 & 0.48958 & 52.1 & 2.46e-5\\
    32 & 0.49479 & 26.0 & 9.83e-5\\
    64 & 0.49740 & 13.0 & 3.93e-4\\
    128 & 0.49870 & 6.51 & 1.57e-3\\
    256 & 0.49935 & 3.26 & 6.29e-3\\
    512 & 0.49967 & 1.63 & 2.52e-2\\
    1024 & 0.49984 & 0.814 & 0.101\\
    2048 & 0.49992 & 0.407 & 0.403\\
    4096 & 0.49996 & 0.204  & 1.61\\
    \vdots & \vdots & \vdots & \vdots\\
    $\infty$ & 0.5 & 0 & $\infty$\\
    \hline
    \end{tabular}
    \caption{The effect of the \texttt{HEALPix} pixelization scheme on a flat horizon map. The temperature residual is a rough estimate of the difference in the Stokes I beam-weighted foreground between the discrete and continuous horizon and was calculated by subtracting the blockage fraction from 0.5 and multiplying by 5,000 K. The map size assumes that the \texttt{HEALPix} map is stored as an array of double-precision values (i.e. \texttt{float64}), the standard data type for floating-point numbers in Python.}
    \label{tab:horizon_nside}
\end{table}

This discretization error can be made arbitrarily small by using maps with higher values of $N_{\textup{side}}$, but large maps quickly become unwieldy due to memory constraints. While the memory required for a given map can be reduced by choosing a data type such as boolean instead of float, using the horizon map for simulations will usually require calculations involving other maps such as the sky temperature and the beam, requiring higher precision. Although we cannot perfectly model the true horizon with a discrete map, we can minimize the error by allowing pixels in the horizon map to take intermediate values within the range $[0,1]$, even when the horizon perfectly blocks incoming radiation. In the case of the flat horizon, we can set the value of the pixels centered at $\theta = 90^{\circ}$ to 0.5, rather than 0 or 1. Although some sources that are below the true horizon will contribute to the discrete calculation of the beam-weighted foreground, this method ensures that the fraction of the sky blocked by the horizon is consistent, i.e.
\begin{equation}
    \frac{\int_{4\pi} \mathcal{H}(\nu, \theta, \phi)d\Omega}{4\pi} = \frac{\sum_{i}^{N_{\textup{pix}}}\mathcal{H}(\nu, i)}{N_{\textup{pix}}}.
\end{equation}
The \texttt{healpy.pixelfunc.ud\_grade} function provides a useful Python function for degrading the resolution of \texttt{HEALPix} maps without changing the fraction of the sky blocked by the horizon. 

\begin{figure*}
    \centering
    \includegraphics[width=\textwidth]{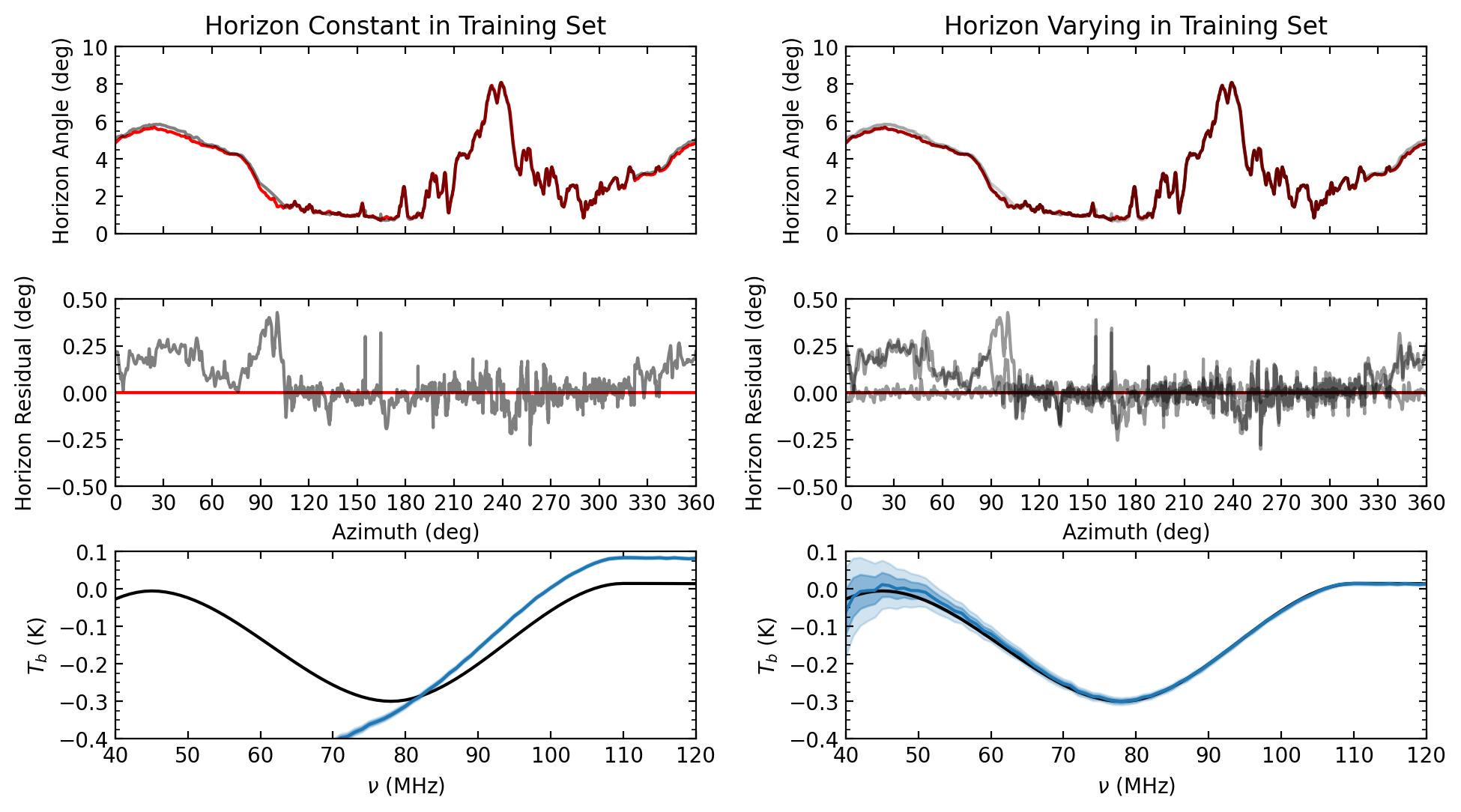}
    \caption{Simulated signal extractions using horizons seen from locations near the center of the Schr\"odinger Basin on the lunar farside. \textit{Top}: Horizon profiles used for the simulated data realization (red) and the foreground training set curves (grey). In the case on the left, only one horizon is included in the training set, whereas on the right, four different horizons are included. \textit{Middle}: Residuals between the horizon profiles used to create the foreground training set (grey) and the fiducial horizon used to create the data realization (red). \textit{Bottom}: 68 and 95\% confidence intervals of the extracted signal compared to the input (black). While the signal extraction is inaccurate when the training set horizon is slightly incorrect (as in the case on the left), including several realizations of the horizon within the training set produces a significant improvement in both the fit and the signal extraction. The $\chi^2_{\textup{red}}$ for the full fit is 220 when the horizon is constant in the training set and 1.02 when the horizon varies within the training set.}
    \label{fig:schrodinger_horizon_uncertainty}
\end{figure*}

To summarize, we propose a simple process for constructing and utilizing a discrete horizon map.
\begin{enumerate}
    \item Create a ``base'' horizon map at the highest possible resolution given the amount of available memory. For this step, utilize the data type that requires the least amount of memory without the loss of information.
    \item Degrade the base map to a lower resolution using \texttt{healpy.pixelfunc.ud\_grade} or a similar function that keeps the sky blockage fraction constant. This lower resolution should be chosen to be the highest resolution such that calculations with this map can be performed efficiently.
\end{enumerate}
It may also be possible to implement a multi-resolution map with adaptive pixel discretization\footnote{\href{https://mhealpy.readthedocs.io/en/latest/index.html}{\texttt{mhealpy}} is one such implementation in Python,\\ \url{https://mhealpy.readthedocs.io/}} to represent the horizon in a more efficient manner, though a more thorough exploration is left to future work. 

\subsection{Accounting for Uncertainty in the Horizon Profile}
\label{accounting_for_uncertainty}

Even if the horizon could be represented exactly in a discrete format, there will always be some level of uncertainty in the horizon profile due to measurement error. One possible source of uncertainty is from error in the elevation measurements used to derive the horizon profile, as discussed in Section \ref{uncertainty_in_horizon_profile}. Another source is uncertainty in the precise location of the antenna. While ground-based experiments should in principle be able to locate the coordinates of the antenna to extreme precision using technologies such as GPS, for illustration purposes we will consider an antenna on the lunar surface (similar in nature to that proposed for DAPPER). Under these circumstances, it is possible there may be uncertainty in the landing location on the order of meters.

We will now demonstrate how our pipeline (as outlined in Papers I-IV) can produce robust signal extractions even when there is significant uncertainty in the shape of the horizon profile. As described in Paper I, we begin the modeling process by creating training sets composed of many realizations of a given component. The foreground training set is composed of simulations of $T_{\textup{ant}}$ (see Equation \ref{continuous_foreground_equation}) which span 10 LST bins and the 4 Stokes polarization parameters (for a total of 40 spectra). While we assume that the beam is given by Equations \ref{beam_equation} and \ref{beam_fwhm_equation}, the foreground training set includes variations in the spatial distribution of the galactic spectral index\footnote{The spectral index is assumed to follow a Gaussian profile centered on the galactic plane with a standard deviation of $10^{\circ}$ as described in Equation 4a of \cite{Hibbard:2020}. Variations are sourced by drawing random values of the offset ($O$) and magnitude ($M$) from the distributions $O \sim \mathcal{N}(-2.5, 0.005)$ and $M \sim \mathcal{N}(-0.4, 0.01)$, where $\mathcal{N}(\mu, \sigma^2)$ indicates the normal distribution with mean $\mu$ and variance $\sigma^2$.} as well as the horizon. For the horizon, we consider an instrument on board a landed payload in the Schr\"odinger Basin on the lunar farside near the coordinates ($75.0^{\circ}$ S, $132.4^{\circ}$ E). We assume that the uncertainty on the exact landing site is 10 meters. To form the training set, we randomly select locations that are Gaussian distributed around the nominal coordinates with a standard deviation of 10 m. We then calculate the horizon at each of these locations via the process described in Section \ref{calculating_horizon} and use the calculated profiles to generate the training set curves. In lieu of a signal training set, we utilize the so-called minimum assumption analysis (MAA) detailed in \cite{Tauscher:2020b}. The MAA assumes only that the 21-cm signal appears the same in all Stokes I spectra, a robust assumption that does not rely on a specific signal model.

The results of fits to a simulated data realization with the horizon from the fiducial coordinates using foreground models derived from two different training sets is shown in Figure \ref{fig:schrodinger_horizon_uncertainty}. On the left, we use only one of the randomly generated horizons to create the training set. As shown in the middle panel, this horizon is slightly different than the one used to create the simulated data realization. In effect, this represents the case in which the horizon is ``lost,'' i.e. the assumed horizon does not match the true horizon. The result of this mismatch is that the signal extraction is extremely poor. The solution to this problem is to include several different horizon realizations in the foreground training set, which span the uncertainty in the horizon profile. As shown in the example in the right panels of Figure \ref{fig:schrodinger_horizon_uncertainty}, the signal extraction is greatly improved by increasing the number of horizons included in the training set. In this case only four horizon profiles are required to produce a good fit to the data and an accurate signal extraction, but more would be needed if the uncertainty on the location of the antenna was significantly larger.

Because of the natural uncertainty in the horizon as a measured quantity, it is crucial for any analysis method to be able to account for this uncertainty. The two examples shown in Figure \ref{fig:schrodinger_horizon_uncertainty} illustrate the power of our training set-based pipeline for accomplishing this task. Although these examples assumed that the uncertainty in the horizon profile propagated from uncertainty in the location of the antenna, other sources of uncertainty such as elevation measurement error can be dealt with in a similar manner. While larger uncertainties may require a greater number of realizations, all that is needed to rigorously account for uncertainty in the horizon with our pipeline is to simulate different profiles that adequately represent the level of uncertainty on the horizon and use them to create the training set curves.

\subsection{Vegetation and Other Foreground Obstructions}
\label{vegetation}

Up to this point, we have only considered a binary horizon profile in which incoming radiation is either completely blocked or entirely unattenuated by the horizon. While this may be true in directions where the horizon profile is defined by large-scale features in the terrain such as hills or mountains, when the horizon is determined by other obstructions such as trees, the effect of the horizon on the propagation of radio waves may change. The main difference is that the sky that is blocked by this intervening media may be partially attenuated without being blocked entirely.

We begin by focusing on vegetation, since it is likely to be the most common of this class of obstructions on Earth. The most common model for expressing the attenuation of a radio wave traveling through foliage is a modified exponential decay model first proposed by \cite{Weissberger:1982}. A more recent version proposed by the radiocommunication sector of the International Telecommunications Union (ITU-R) is given by
\begin{equation}
\label{vegetation_attenuation_equation}
    L = Af^Bd^C(\theta + E)^G,
\end{equation}
where $f$ is the frequency in MHz, $d$ is the path length\footnote{Assuming a rectangular ``block'' of vegetation, the path length can be calculated by $d = h\big(1 - \frac{\tan \theta}{\tan \eta}\big)\csc \theta$, where $h$ is the vertical height of the canopy, $\eta$ is the elevation angle of the top of the canopy from the observer, and $\theta$ is the elevation angle along the line-of-sight through the vegetation.} in meters, $\theta$ is the elevation angle in degrees, and $A$, $B$, $C$, $E$, and $G$ are parameters that are generally determined empirically and may be influenced by the physical properties of the vegetation such as leaf size. Assuming the validity of this model, the values of the empirically determined parameters are likely to be highly dependent on the local flora as well as possibly on time of year. Thus, for experiments in locations where trees or other vegetation define the horizon, there may be significant uncertainty in the parameter values. In the following example, we show how our pipeline can account for attenuation due to vegetation even when the exact values of the parameters of the attenuation model are not known.

\begin{figure}
    \centering
    \includegraphics[width=\columnwidth]{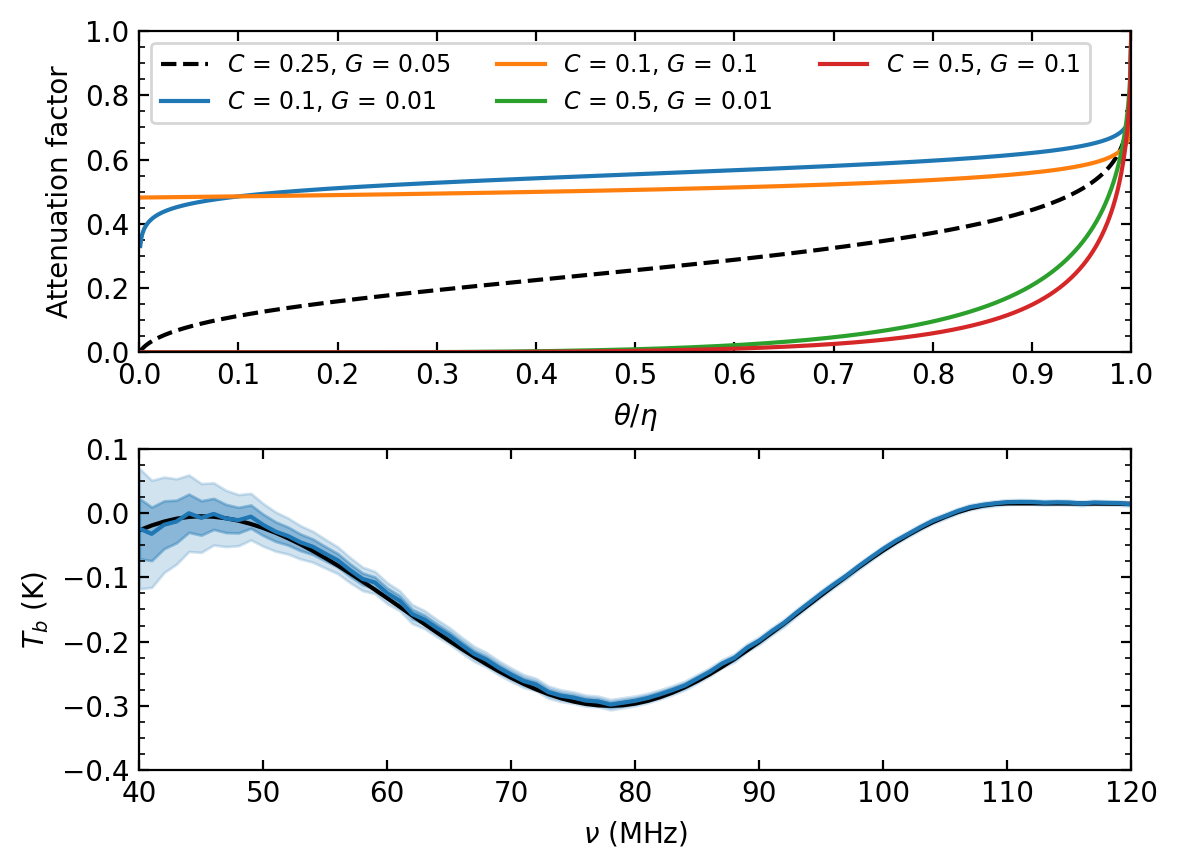}
    \caption{\textit{Top}: Attenuation due to vegetation as a function of the ratio of the elevation angle of the line-of-sight through the vegetation ($\theta$) to the elevation angle of the top of the canopy ($\eta$) for the model in Equation \ref{vegetation_attenuation_equation}. The dashed black curve indicates the attenuation used to create the data realization, while the solid colored curves indicate the levels of attenuation used to construct the foreground training set. \textit{Bottom}: Simulated signal extraction with the data realization and foreground training sets using the attenuation levels described in the top panel. The contours indicate the 68 and 95\% confidence intervals while the black curve indicates the input signal. The $\chi^2_{\textup{red}}$ of the full fit is 0.99 and the signal is extracted accurately.}
    \label{fig:vegetation_loss_model}
\end{figure}

To simulate a horizon that includes vegetation, we utilize the horizon from GBO shown in Figure \ref{fig:greenbank_horizon}. As discussed in Section \ref{comparison_to_measurement}, the direct measurements follow the top of the tree canopy, while the horizon calculated from elevation data does not take the trees into account. Thus, we can create a horizon map in which the sky is completely blocked below the calculated horizon, but for points between the calculated and measured horizons, the sky is attenuated according to the model in Equation \ref{vegetation_attenuation_equation}.

To create a simulated data realization, we will use the best fit values provided in the ITU-R attenuation in vegetation report, $A = 0.25$, $B = 0.39$, $C = 0.25$, $E = 0$, and $G = 0.05$.\footnote{These values were inferred from measurements made in pine woodland in Austria. While the trees surrounding GBO are likely to change these values, we are simply using them as an example rather than trying to rigorously model the conditions at Green Bank.} For simplicity, in this example we will ignore the frequency dependence, fixing $f = 100$ MHz for the purposes of the attenuation model. In attempting to model the foreground, we will act under the assumption that the values of $C$ and $G$ are not well constrained. To create the foreground training set, we will use $C = [0.1, 0.5]$ and $G = [0.01, 0.1]$. The top panel of Figure \ref{fig:vegetation_loss_model} shows the attenuation as a function of elevation angle using these values for $C$ and $G$. Note that while the training set does not contain any realizations made using the true values of $C$ and $G$, the values included in the training set encompass the true values. The bottom panel of Figure \ref{fig:vegetation_loss_model} shows the simulated signal extraction. In a similar manner to the example in Section \ref{accounting_for_uncertainty} with the shape of the horizon profile changing with the observer's location, the power of the training set-based method is that we do not need to know the exact values of the parameters of the vegetation attenuation model, we only need to include values for the parameters in the training set that encompass the true values. In this example we assumed that the attenuation was cause by vegetation, but this framework can be extended to other obstructions such as human-made objects that may be near an instrument, so long as there is a reasonable model of the attenuation due to the object. While here we have ignored any spectral variation in the attenuation for the purposes of this example, further tests have shown that uncertainty in the value of $B$ (the exponent of frequency in Equation 
\ref{vegetation_attenuation_equation}) can be dealt with in a similar manner.

\section{Conclusions}
\label{conclusions}

The primary goal of this work is to demonstrate that the horizon has a significant impact on the beam-weighted foreground component of low frequency observations. Although the difference in the beam-weighted foreground imposed by a given horizon is generally smooth and power law-like in frequency, the spatial anisotropy of the foreground emission and apparent movement of the Galaxy across the sky introduce a complex time dependence. While the exact time dependence is determined by the shape of the horizon profile and the latitude of the observer, in general the effect of the horizon will be maximized when the plane of the Galaxy is near the horizon and minimized when the Galaxy is highest in the sky or furthest beneath the horizon.

We also attempt to explain the importance of the horizon in modeling observational data. Without knowledge of the horizon, one cannot forward model the beam-weighted foreground component, even if the sky map and the antenna beam are known exactly. Thus, in terms of accurately modeling observations, the horizon cannot be ignored.

Beyond simply modeling data, the horizon may in some cases influence fits that attempt to extract the 21-cm global signal. We find that in a single spectrum fit (e.g., see \citealt{Bowman:2018}), the bias imposed by the horizon can often be absorbed by the foreground model, meaning that precise knowledge of the horizon may not be necessary. When fitting multiple spectra, such as when data are binned in LST or when measurements include the four Stokes polarization parameters, the foreground model can again generally absorb the horizon bias, but only if the spectra are treated independently, meaning that there are separate parameters for each spectrum. The problem with both of these methods of analysis is that they do not provide adequate constraints on the 21-cm signal. Through simulations, signal extractions produced by single spectrum fits were found to have RMS uncertainties at the Kelvin level and greater (for a thorough explanation, see Paper III). In order to reduce the uncertainty in the signal extraction, multiple spectra cannot be treated independently but instead must be correlated. In other words, only one set of parameters is used for all spectra. When the spectra are correlated, however, we naturally find that much more precise knowledge of the horizon is required to fit the data well. Although performing a correlated fit requires more detailed modeling, Paper III shows that it is necessary to produce precise constraints on the signal.

After illustrating the importance of the horizon for global 21-cm analysis, we demonstrate how to model it to a sufficient precision to extract the global signal in correlated, multi-spectrum fits. Our pipeline, which relies on training sets that include many simulated realizations of each component, can still robustly model the beam-weighted foreground even in the presence of significant uncertainty in various properties of the horizon. In order to be able to include the horizon in the simulated realizations that make up the foreground training set, we derive an algorithm and present an open source Python package for calculating the horizon profile at a given location based on elevation data. Adequately modeling the horizon, however, requires additional details beyond just the shape of the horizon profile obtained from elevation data. While large scale structures in the terrain such as hills or mountains may block incoming radiation entirely, vegetation and other foreground obstructions may only partially attenuate radio waves. Luckily, the attenuation of a radio wave traveling through vegetation has been found to follow a relatively simple exponential decay model. Although the parameters of the model may vary from location to location (or even season to season) based on the properties of the foliage, we find that the pipeline can still produce an adequate foreground model if several realizations of the attenuation model are included in the training set with parameter values that encompass the true values. We also consider the effect of uncertainty (from either uncertainty in the location of the antenna or measurement error in the elevation data) in the shape of the horizon. Again, we find that our pipeline can handle this uncertainty by including realizations of the horizon in the training set that vary according to the level of uncertainty.

After a careful consideration, it is clear that the problem of the ``lost horizon'' cannot be ignored when performing precision global 21-cm analysis. Although in principle the horizon at any location can be modeled to a sufficient level to extract the 21-cm signal precisely, in reality the horizon presents a daunting challenge for forward modeling-based analysis methods. However, the problem of modeling the horizon can be simplified by choosing an optimal location for performing observations. While we have presented examples of how to potentially model features such as vegetation, the ideal horizon would be one that is free of obstructions that may interact with incoming radiation in complex, and perhaps unknown, ways. We conclude by noting that one such location that is entirely free of vegetation of any kind is the surface of the Moon. In addition to the radio quiet farside and the lack of a significant ionosphere, the simplicity of the horizon provides yet another reason why the Moon may be Shangri-La for 21-cm experiments.

\section{Acknowledgements}

This work is directly supported by the NASA Solar System Exploration Research Virtual Institute cooperative agreement number 80ARC017M0006. This work was also supported by the National Aeronautics and Space Administration (NASA) under award number NNA16BD14C for NASA Academic Mission Services. BN is a Jansky Fellow of the National Radio Astronomy Observatory (NRAO). DB is a Grote Reber Doctoral Fellow at NRAO.

\bibliographystyle{aasjournal}
\bibliography{ref}

\appendix
\section{Extracting Horizon from Elevation Data}
\label{horizon_derivation}

\begin{figure*}
    \centering
    \includegraphics[width=\textwidth]{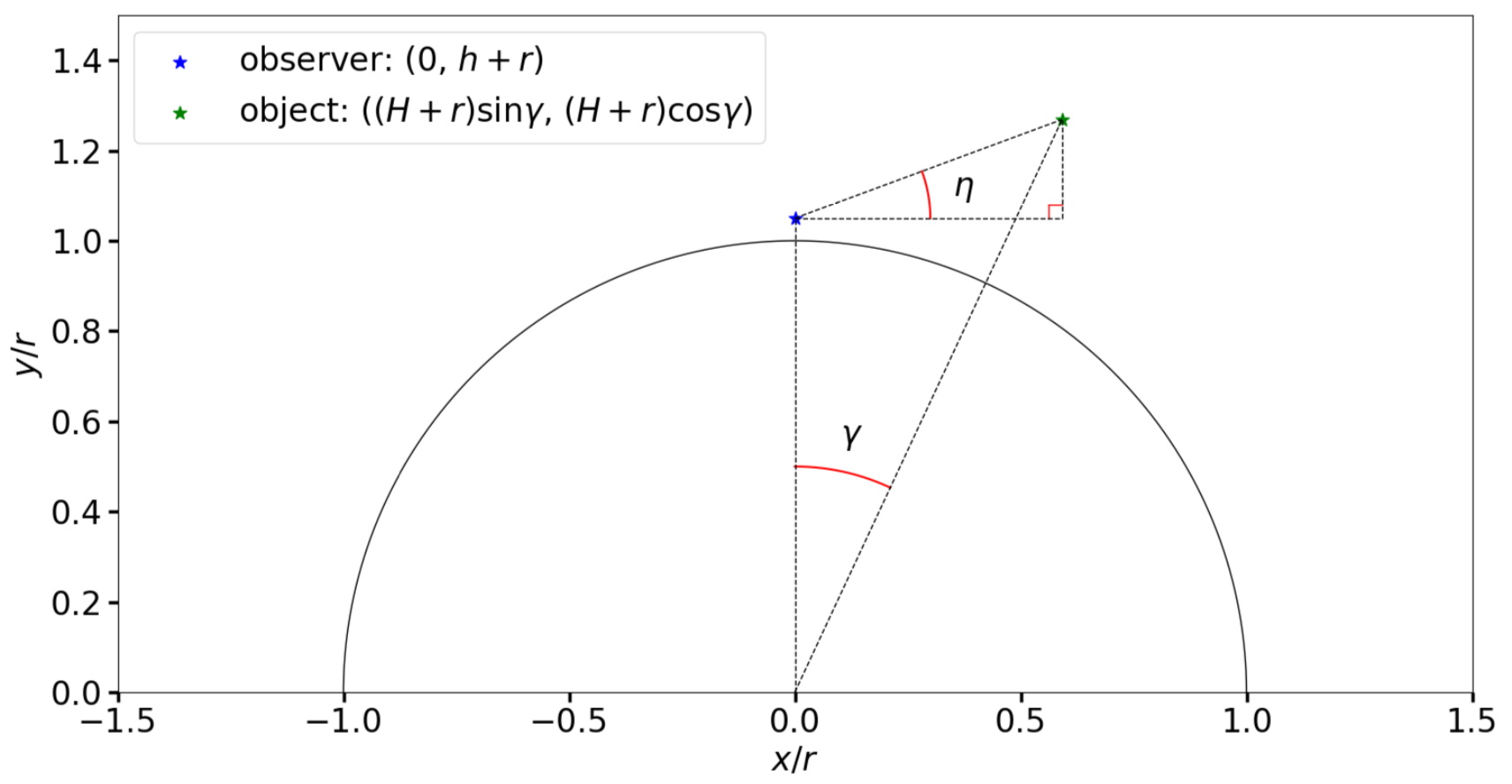}
    \caption{Schematic showing the positions of the observer and object with respect to the body along with the angles defined in Appendix \ref{horizon_derivation}.}
    \label{fig:horizon_diagram}
\end{figure*}

This appendix presents an algorithm for extracting the angular horizon from any given location on a spherical body using elevation data. Throughout the following derivation, $r$ represents the radius of a (roughly) spherical body and $h$ the elevation of an observer on its surface. First, we will derive an equation that yields the angular height of an object an angular distance $\gamma$ away on the body that is at an elevation of $H$. Next, we illustrate how to find the horizon angle in a constant direction. Finally, we show how to parameterize the great circle corresponding to a given direction so that the 1D elevation profile of the positions in a single direction can be effectively queried from data.

We orient the body so that the plane containing the center of the body, the observer, and the object is the $xy$-plane. We place the center of the body at the origin and the observer at ($x_{\textup{obs}}$, $y_{\textup{obs}}$) = (0, $r+h$). We also orient the plane so that the object has a positive $x$-coordinate. If the angular distance to the object (measured from the center of the body) is $\gamma$ and its height is $H$, the object is located at ($x_{\textup{obj}}$, $y_{\textup{obj}}$) = ($(r+H)\sin{\gamma}$, $(r+H)\cos{\gamma}$). The tangent of the angle from the horizontal to the object (up being positive) is given
by
\begin{equation}
    \tan{\eta} = \frac{y_{\textup{obj}} - y_{\textup{obs}}}{x_{\textup{obj}} - x_{\textup{obs}}}.
\end{equation}
Plugging in the expressions above, we find
\begin{equation}
    \tan{\eta} = \frac{(r+H)\cos{\gamma} - (r + h)}{(r+H)\sin{\gamma}},
\end{equation}
which is the same as
\begin{equation}
    \tan{\eta} = \cot{\gamma} - \Big(\frac{r+h}{r+H}\Big)\csc{\gamma}.
\end{equation}
Since $|\eta| \leq \pi / 2$, taking the arctangent of both sides yields
\begin{equation}
\label{eta_equation}
    \eta = \arctan \Big[\cot{\gamma} - \Big(\frac{r+h}{r+H}\Big)\csc{\gamma}\Big].
\end{equation}

Equation \ref{eta_equation} gives the angle from the horizontal of a given object at a constant elevation $H$ and angular distance $\gamma$ from the observer. The horizon angle $\beta$ in a given direction is the maximum value of $\eta$ in that direction as a function of $\gamma$, i.e.
\begin{equation}
\label{beta_equation}
    \beta = \max_{\gamma \in [0, \gamma_{\textup{max}}]} \arctan \Big[\cot{\gamma} - \Big(\frac{r+h(0)}{r+h(\gamma)}\Big)\csc{\gamma}\Big],
\end{equation}
where $h(\gamma)$ is the elevation profile as a function of angular distance. Here, $\gamma_{\textup{max}}$ is the maximum angular distance considered and can be written as
\begin{equation}
\label{gamma_max_equation}
    \gamma_{\textup{max}} = \arccos \Big(\frac{r + h_{\textup{min}}}{r + h_{\textup{max}}}\Big),
\end{equation}
where $h_{\textup{min}}$ and $h_{\textup{max}}$ are the minimum and maximum elevations on the body under consideration. At angular distances above $\gamma_{\textup{max}}$, the highest elevation on the body would appear below the horizontal from the lowest elevation on the body.\footnote{On Earth, taking $r = 6300$ km, $h_{\textup{min}} = −500$ m (Dead Sea depression), and $h_{\textup{max}} = 9000$ m (Mount Everest peak), we have $\gamma_{\textup{max}} \approx 3.2^{\circ}$. On the Moon, taking $r = 1700$ km, $h_{\textup{min}} = −9000$ m (center of inner crater of Antoniadi Crater in the South Pole-Aitken basin), and $h_{\textup{max}} = 5500$ m (Mount Huygens peak), we have $\gamma_{\textup{max}} \approx 7.5^{\circ}$. In practice, $\gamma_{\textup{max}}$ can be calculated using a local, rather than absolute, minimum and maximum.}

We can parameterize any great circle traveling through the observer’s position as
\begin{equation}
    \hat{\boldsymbol{n}}(\gamma) = \hat{\boldsymbol{r}}_0\cos{\gamma} + (\hat{\boldsymbol{\phi}}_0\sin{\alpha} - \hat{\boldsymbol{\theta}}_0\cos{\alpha})\sin{\gamma},
\end{equation}
where $\hat{\boldsymbol{n}}$ is the normal vector and $\alpha$ is the azimuthal angle the great circle describes from the starting point $\hat{\boldsymbol{r}}_0$, which points toward the observer's position ($\gamma = 0$) from the center of the body, with $\alpha=0$ being north and $\alpha=\frac{\pi}{2}$ being east. $\hat{\boldsymbol{r}}_0$, $\hat{\boldsymbol{\theta}}_0$, and $\hat{\boldsymbol{\phi}}_0$ are the unit vectors in the radial, polar, and azimuthal directions at the observer's position. This implies that the polar and azimuthal angles as a function of $\gamma$ are
\begin{subequations}
    \begin{equation}
    \label{theta_equation}
        \theta (\gamma) = \arccos (\cos \theta_0 \cos \gamma + \cos \alpha \sin \theta_0 \sin \gamma),
    \end{equation}
    \begin{align}
    \label{phi_equation}
        \phi (\gamma) = \arg \big[(\sin \theta_0 \cos \gamma + i \sin \alpha \sin \gamma - \cos \alpha \cos \theta_0 \sin \gamma)e^{i\phi_0}\big],
    \end{align}
\end{subequations}
where $\arccos$ is the inverse cosine function (returning an angle in radians) and $\arg$ is the function that returns the argument of a complex number between $-\pi$ and $\pi$ radians. The latitude, $b$, and longitude, $\lambda$, are defined from $\theta$ and $\phi$ through
\begin{subequations}
    \begin{equation}
    \label{b_equation}
        b(\gamma) = 90^{\circ} - \Big(\frac{180^{\circ}}{\pi}\Big)\theta(\gamma),
    \end{equation}
    \begin{equation}
    \label{lambda_equation}
        \lambda (\gamma) = \Big(\frac{180^{\circ}}{\pi}\Big)\phi (\gamma).
    \end{equation}
\end{subequations}
Appendix \ref{algorithm_pseudocode_appendix} presents pseudo-code that implements the equations derived in this section for calculating the horizon profile.

\newpage
\section{Horizon Profile Algorithm}
\label{algorithm_pseudocode_appendix}

This appendix presents an example of a pseudo-code definition of two functions that will compute the horizon profile given the equations derived in Appendix \ref{horizon_derivation}.
\begin{algorithmic}[1]
    \Function{horizon\_angle}{$\alpha, \gamma$}
        \State $\theta(\gamma) \leftarrow$ Equation \ref{theta_equation}
        \State $\phi(\gamma) \leftarrow$ Equation \ref{phi_equation}
        \State $b(\gamma) \leftarrow$ Equation \ref{b_equation}
        \State $\lambda(\gamma) \leftarrow$ Equation \ref{lambda_equation}
        \State Compute $h(\gamma)$ from elevation data queries at $b(\gamma)$ latitude and $\lambda(\gamma)$ longitude
        \State $\eta \leftarrow$ Equation \ref{eta_equation}
        \State \Return $\eta$
    \EndFunction
    
    \Function{horizon\_profile}{$N_{\alpha}, N_{\gamma}$}
        \State azimuths $\leftarrow$ linspace(0, $2\pi$, $N_{\alpha}$)
        \State $\gamma_{\textup{max}} \leftarrow$ Equation \ref{gamma_max_equation}
        \State gammas $\leftarrow$ linspace(0, $\gamma_{\textup{max}}$, $N_{\gamma}$)
        \State horizon\_profile $\leftarrow$ [ ]
        \For{$\alpha$ in azimuths}
            \State horizon\_angles $\leftarrow$ [ ]
            \For{$\gamma$ in gammas}
                \State $\eta \leftarrow$ {\footnotesize HORIZON\_ANGLE}($\alpha, \gamma$)
                \State Append $\eta$ to horizon\_angles
            \EndFor
            \State $\beta$ $\leftarrow$ maximum(horizon\_angles)
            \State Append $\beta$ to horizon\_profile
        \EndFor
        \State \Return azimuths, horizon\_profile
    \EndFunction
\end{algorithmic}

\section{Horizon Profile Code}
\label{horizon_profile_code}

An implementation of the algorithm presented in Appendix \ref{algorithm_pseudocode_appendix} is made publicly available through the Simulating Horizon Angle Profile from Elevation Sets (SHAPES) code.\footnote{\url{https://github.com/npbassett/shapes}} The \texttt{shapes} code provides the ability to calculate the horizon for most locations on both the Earth and the Moon, through data provided by either the Shuttle Radar Topography Mission \citep[SRTM]{Farr:2007} or the Lunar Orbiter Laser Altimeter \citep[LOLA]{Smith:2010}.

The SRTM data is accessed through the \texttt{elevation} Python package.\footnote{\url{https://pypi.org/project/elevation/}} The SRTM 1 arcsecond dataset has an absolute accuracy of 16 meters at 90\% confidence. While it is known that the accuracy decreases with increases in slope and elevation, such as in the Himalayas where large outliers and measurement voids exist \citep{Mukul:2017}, areas surrounding most global signal instruments are unlikely to be rugged enough for this to be an issue. Lunar calculations employ the LOLA 118m dataset,\footnote{\url{https://astrogeology.usgs.gov/search/details/Moon/LRO/LOLA/Lunar_LRO_LOLA_Global_LDEM_118m_Mar2014/cub}} which covers the entire lunar surface with a vertical accuracy of $\sim$ 1 m \citep{Mazarico:2012}. For both data sets, a 2D linear interpolation is used in order to calculate elevations between data points.

\section{Horizon Comparison}
\label{horizon_comparison}

\begin{figure*}[h]
    \centering
    \includegraphics[width=0.9\textwidth]{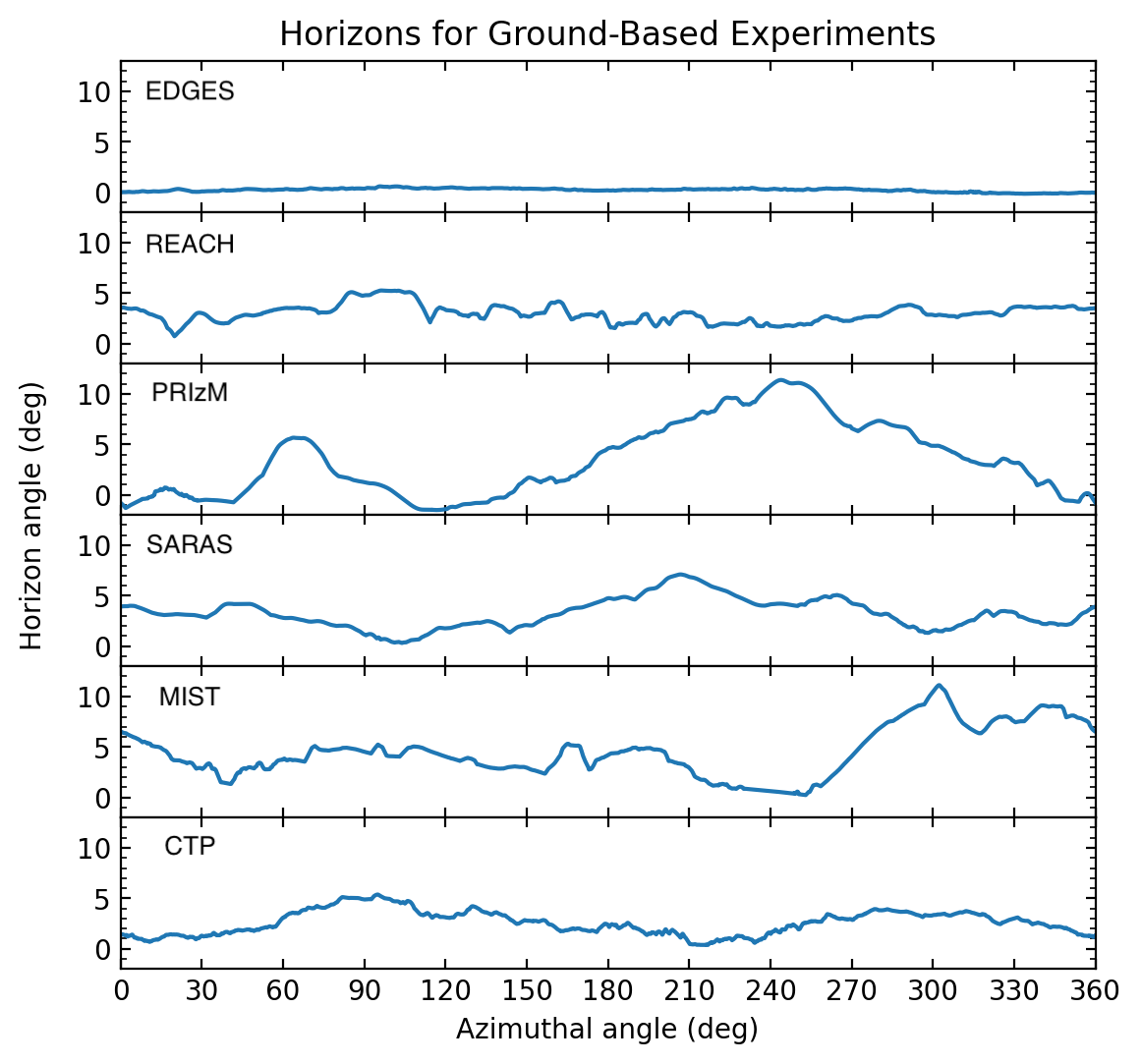}
    \caption{Horizon profiles calculated by the \texttt{shapes} code for a selection of different global 21-cm experiments.}
    \label{fig:experiment_horizons}
\end{figure*}

For comparison and reference purposes, we have calculated the apparent horizon for a handful of global 21-cm experiments, which are shown in Figure \ref{fig:experiment_horizons}: the Experiment to Detect the Global EoR Signal (EDGES) in the Murchison radio observatory in Western Australia, the Radio Experiment for the Analysis of Cosmic Hydrogen (REACH) in the Karoo radio reserve in South Africa, Probing Radio Intensity at High-z from Marion (PRIZM) on Marion Island in the Subantarctic Indian Ocean, the Shaped Antenna measurement of the background RAdio Spectrum (SARAS) in the Gauribidanur Radio Observatory in India, the Mapper of the IGM Spin Temperature (MIST) at McGill Arctic Research Station in Northern Canada, and the Cosmic Twilight Polarimeter (CTP) in Green Bank Observatory in West Virginia, United States. Note that, as mentioned in Section \ref{comparison_to_measurement}, these calculations may not reflect foreground obstructions such as vegetation. While we attempted to pinpoint the locations as precisely as possible, the coordinates used to calculate the horizon may not reflect the true locations of the instruments. The coordinates used in the calculations were:
\begin{itemize}
    \item EDGES: \href{https://goo.gl/maps/u8s7owniWV8TYUK38}{$26^{\circ}\ 42^{\prime}\ 53.8^{\prime\prime}$ S, $116^{\circ}\ 36^{\prime}\ 12.5^{\prime\prime}$ E}
    \item REACH: \href{https://goo.gl/maps/iy2ic9qbDd7URzFCA}{$30^{\circ}\ 50^{\prime}\ 01.0^{\prime\prime}$ S, $21^{\circ}\ 22^{\prime}\ 33.0^{\prime\prime}$ E}
    \item PRIzM: \href{https://goo.gl/maps/vkaw8qhBpBBwmkre8}{$46^{\circ}\ 53^{\prime}\ 12.6^{\prime\prime}$ S, $37^{\circ}\ 49^{\prime}\ 12.2^{\prime\prime}$ E}
    \item SARAS: \href{https://goo.gl/maps/QFwEU1oC7GqKSmvU7}{$13^{\circ}\ 59^{\prime}\ 31.8^{\prime\prime}$ N, $74^{\circ}\ 52^{\prime}\ 33.7^{\prime\prime}$ E}
    \item MIST: \href{https://goo.gl/maps/6U3LRuH5rKm8Fwmg6}{$79^{\circ}\ 24^{\prime}\ 54.7^{\prime\prime}$ N, $90^{\circ}\ 44^{\prime}\ 56.2^{\prime\prime}$ W}
    \item CTP: \href{https://goo.gl/maps/eERmBVrQcRujUc3Q6}{$38^{\circ}\ 26^{\prime}\ 1.4^{\prime\prime}$ N, $79^{\circ}\ 49^{\prime}\ 22.9^{\prime\prime}$ W}
\end{itemize}

\section{Error in Horizon Profile}
\label{horizon_error_appendix}

\subsection{Error in \texorpdfstring{$\gamma$}{gamma}}

The elevation data used in creating horizon profiles pixelates the surface of the body being mapped, leading to a possible error in the angle $\gamma$ of $\pm \delta \gamma$. To find an estimate of $\delta\gamma$, we find the angle between two opposite corners of a pixel, one at $(\theta, \phi)$ and one at $(\theta+\delta\theta,\phi+\delta\phi)$, where $\delta\theta$ and $\delta\phi$ are the resolutions in $\theta$ and $\phi$, respectively. $\delta\gamma$ satisfies
\begin{equation}
  \cos{(\delta\gamma)} = \cos{(\theta)}\cos{(\theta+\delta\theta)} + \sin{(\theta)}\sin{(\theta+\delta\theta)}\cos{(\delta\phi)}.
\end{equation}
Using the angle sum and difference formulas for sine and cosine, this can be written
\begin{equation}
  (1 - \cos{\delta\gamma}) = (1 - \cos{\delta\theta}) + (1 - \cos{\delta\phi})\sin{\theta}(\sin{\theta}\cos{\delta\theta} + \cos{\theta}\sin{\delta\theta}).
\end{equation}
To lowest order in the resolution parameter, i.e. assuming $1-\cos{\delta\gamma}=(\delta\gamma)^2/2$, $1-\cos{\delta\theta}=(\delta\theta)^2/2$, $1-\cos{\delta\phi}=(\delta\phi)^2/2$, and $\sin{\theta}\cos{\delta\theta} + \cos{\theta}\sin{\delta\theta}=\sin{\theta}$,\footnote{These basically amount to assuming that $\delta\gamma$, $\delta\theta$, and $\delta\phi$ are much less than one in radians, which is justified for maps that could be used to produce reasonable horizon profiles.} we find
\begin{equation}
  \delta\gamma = \sqrt{(\delta\theta)^2+(\delta\phi)^2\sin^2{\theta}}.
\end{equation}
Since the latitude $b$ is $(\pi/2)-\theta$ and the longitude $\lambda$ is $\phi$,
\begin{equation}
  \delta\gamma = \sqrt{(\delta b)^2+(\delta\lambda)^2\cos^2{b}}.
\end{equation}

\subsection{Errors in \texorpdfstring{$h$}{h} and \texorpdfstring{$H$}{H}}

We assume that the errors in the elevation map are correlated, satisfying
\begin{subequations}
\begin{align}
  \textup{Var}[h] &= (\delta h)^2, \\
  \textup{Var}[H] &= (\delta H)^2, \\
  \textup{Cov}[h,H] &= \rho(\gamma)(\delta h)(\delta H).
\end{align}
\end{subequations}

\subsection{Derivatives of Horizon Angle}

The derivatives of the tangent of the horizon angle with respect to the parameters $\gamma$, $h$, and $H$ are
\begin{subequations}
\begin{align}
  \frac{\partial\tan{\eta}}{\partial\gamma} &= -\csc^2{\gamma}\left[1 - \left(\frac{r+h}{r+H}\right)\cos{\gamma}\right],\\
  \frac{\partial\tan{\eta}}{\partial h} &= -\csc{\gamma}\left(\frac{1}{r+H}\right),\\
  \frac{\partial\tan{\eta}}{\partial H} &= \csc{\gamma}\left(\frac{r+h}{r+H}\right)\left(\frac{1}{r+H}\right).
\end{align}
\label{eq:tangent_derivatives}
\end{subequations}

\subsection{Error in Horizon Angle}

Assuming that the errors do not cause the wrong object to be determined to create the highest elevation line of sight at a given azimuth, the variance of the tangent of the horizon angle is
\begin{multline}
  \Var[\tan{\eta}] = \left(\frac{\partial\tan{\eta}}{\partial\gamma}\right)^2(\delta\gamma)^2 + \left(\frac{\partial\tan{\eta}}{\partial h}\right)^2(\delta h)^2 + \left(\frac{\partial\tan{\eta}}{\partial H}\right)^2(\delta H)^2 \\ + 2\rho(\gamma)(\delta h)\ (\delta H)\left(\frac{\partial\tan{\eta}}{\partial h}\right)\left(\frac{\partial\tan{\eta}}{\partial H}\right).
\end{multline}
Plugging in Equations \ref{eq:tangent_derivatives}, this becomes
\begin{multline}
  \Var[\tan{\eta}] = \csc^4{\gamma}\left[1 - \left(\frac{r+h}{r+H}\right)\cos{\gamma}\right]^2\left[(\delta b)^2+(\delta\lambda)^2\cos^2{b}\right] \\ + \csc^2{\gamma}\left(\frac{\delta h}{r+H}\right)^2 + \csc^2{\gamma}\left(\frac{r+h}{r+H}\right)^2\left(\frac{\delta H}{r+H}\right)^2 \\ - 2\rho(\gamma)\csc^2{\gamma}\left(\frac{r+h}{r+H}\right)\left(\frac{\delta h}{r+H}\right)\left(\frac{\delta H}{r+H}\right).
\end{multline}
Reordering, we obtain
\begin{multline}
  \Var[\tan{\eta}] = \csc^4{\gamma}\left[1 - \left(\frac{1+(h/r)}{1+(H/r)}\right)\cos{\gamma}\right]^2\left[(\delta b)^2+(\delta\lambda)^2\cos^2{b}\right] \\ + \csc^2{\gamma}\left(\frac{(\delta h/r)}{1+(H/r)}\right)^2 + \csc^2{\gamma}\left(\frac{1+(h/r)}{1+(H/r)}\right)^2\left(\frac{(\delta H/r)}{1+(H/r)}\right)^2 \\ - 2\rho(\gamma)\csc^2{\gamma}\left(\frac{1+(h/r)}{1+(H/r)}\right)\left(\frac{(\delta h/r)}{1+(H/r)}\right)\left(\frac{(\delta H/r)}{1+(H/r)}\right).
\end{multline}
Since $\Var[\tan{\eta}]=\sec^4{\eta}\ \Var[\eta]$, this means
\begin{multline}
  \Var[\eta] = \cos^4{\eta}\csc^4{\gamma}\left[1 - \left(\frac{1+(h/r)}{1+(H/r)}\right)\cos{\gamma}\right]^2\left[(\delta b)^2+(\delta\lambda)^2\cos^2{b}\right] \\ + \cos^4{\eta}\csc^2{\gamma}\left(\frac{(\delta h/r)}{1+(H/r)}\right)^2 + \cos^4{\eta}\csc^2{\gamma}\left(\frac{1+(h/r)}{1+(H/r)}\right)^2\left(\frac{(\delta H/r)}{1+(H/r)}\right)^2 \\ - 2\rho(\gamma)\cos^4{\eta}\ \csc^2{\gamma}\left(\frac{1+(h/r)}{1+(H/r)}\right)\left(\frac{(\delta h/r)}{1+(H/r)}\right)\left(\frac{(\delta H/r)}{1+(H/r)}\right).
\end{multline}
If the correlation between variations in $h$ and $H$ is given by $\rho(\gamma)=e^{-\gamma/\zeta}$, where $\zeta$ is the correlation length, then
\begin{multline}
  \Var[\eta] = \cos^4{\eta}\ \csc^4{\gamma}\ \left[1 - \left(\frac{1+(h/r)}{1+(H/r)}\right)\cos{\gamma}\right]^2\ \left[(\delta b)^2+(\delta\lambda)^2\cos^2{b}\right] \\ + \cos^4{\eta}\ \csc^2{\gamma}\ \left(\frac{(\delta h/r)}{1+(H/r)}\right)^2 + \cos^4{\eta}\ \csc^2{\gamma}\ \left(\frac{1+(h/r)}{1+(H/r)}\right)^2\ \left(\frac{(\delta H/r)}{1+(H/r)}\right)^2 \\ - 2\ e^{-\gamma/\zeta}\ \cos^4{\eta}\ \csc^2{\gamma}\ \left(\frac{1+(h/r)}{1+(H/r)}\right)\ \left(\frac{(\delta h/r)}{1+(H/r)}\right)\ \left(\frac{(\delta H/r)}{1+(H/r)}\right).
\end{multline}

\end{document}